\documentclass[12pt]{article}
\usepackage{a4wide}
\usepackage{amssymb,amsmath}
\usepackage{graphicx}

\usepackage{subcaption}
\usepackage{tensor}

\begin{document}
{\renewcommand{\thefootnote}{\fnsymbol{footnote}}
\begin{center}
{\LARGE  Emergent modified gravity:\\ The perfect fluid and gravitational collapse}\\
\vspace{1.5em}
Erick I.\ Duque\footnote{e-mail address: {\tt eqd5272@psu.edu}}
\\
\vspace{0.5em}
Institute for Gravitation and the Cosmos,\\
The Pennsylvania State
University,\\
104 Davey Lab, University Park, PA 16802, USA\\
\vspace{1.5em}
\end{center}
}

\setcounter{footnote}{0}

\begin{abstract}
Emergent modified gravity is a canonical theory based on general covariance where the spacetime is not fundamental, but rather an emergent object. This feature allows for modifications of the classical theory and can be used to model new effects, such as those suggested by quantum gravity. We discuss how matter fields can be coupled to emergent modified gravity, realize the coupling of the perfect fluid, identify the symmetries of the system, and explicitly obtain the Hamiltonian in spherical symmetry. We formulate the Oppenheimer-Snyder collapse model in canonical terms, permitting us to extend the model to emergent modified gravity and obtain an exact solution to the dust collapsing from spatial infinity including some effects suggested by quantum gravity. In this solution the collapsing dust forms a black hole, then the star radius reaches a minimum with vanishing velocity and finite positive acceleration, and proceeds to emerge out now behaving as a white hole. While the geometry on the minimum-radius surface is regular in the vacuum, it is singular in the presence of dust. However, the fact that the geometry is emergent, and the fundamental fields that compose the phase-space are regular, allows us to continue the canonical solution in a meaningful way, obtaining the global structure for the interior of the star. The star-interior solution is complemented by the vacuum solution describing the star-exterior region by a continuous junction at the star radius. This gluing process can be viewed as the imposition of boundary conditions, which is non-unique and does not follow from the equations of motion. This ambiguity gives rise to different possible physical outcomes of the collapse. We discuss two such phenomena: the formation of a wormhole and the transition from a black hole to a white hole.
\end{abstract}

\section{Introduction}

General relativity (GR), in its canonical formulation, is a constrained gauge field theory where the gauge transformations, which are generated by the Hamiltonian constraint and the diffeomorphism constraint, are equivalent to spacetime coordinate transformations only on-shell, that is, when the gauge generators vanish.
The spacetime is foliated by space-like hypersurfaces, and the field content is given by a set of spatial tensors on these hypersurfaces, with flow equations generated by the same constraints determining the evolution of these fields between adjacent hypersurfaces.
These evolving spatial tensors can then reproduce the usual spacetime tensors of GR.
The action of the gauge generators can also be understood geometrically: the Hamiltonian constraint $H[N]$, smeared by a scalar lapse function $N$, generates a normal, infinitesimal hypersurface deformation with length $N$, while the diffeomorphism constraint $\Vec{H}[\Vec{N}]$, smeared by the spatial shift vector $\Vec{N}$, generates a tangential hypersurface deformation with length $\Vec{N}$.

In ADM notation \cite{ADM,arnowitt2008republication}, the diffeomorphism constraint $\Vec{H}[\Vec{N}]$ and the Hamiltonian constraint have Poisson brackets
\begin{eqnarray}
    \{ \Vec{H} [ \Vec{N}_1] , \Vec{H} [ \Vec{N}_2 ] \} &=& - \vec{H} [\mathcal{L}_{\Vec{N}_2} \Vec{N}_1]
    \ ,
    \label{eq:Hypersurface deformation algebra - HaHa}
    \\
    \{ H [ N ] , \Vec{H} [ \Vec{N}]\} &=& - H [ N^b \partial_b N ]
    \label{eq:Hypersurface deformation algebra - HHa}
    \ , \\
    \{ H [ N_1 ] , H [ N_2 ] \} &=& - \vec{H} [ q^{a b} ( N_2 \partial_b N_1 - N_1 \partial_b N_2 )]
    \ ,
    \label{eq:Hypersurface deformation algebra - HH}
\end{eqnarray}
that depend not only on $\Vec{N}$ and $N$, but also on the inverse of the spatial metric $q_{a b}$ on a spatial hypersurface.
When a phase-space function, such as $q^{a b}$, appears in the argument of the constraint algebra, it is called a structure function, in contrast to the structure constants of usual Lie algebras.
This constraint algebra, also known as the hypersurface deformation algebra, is central to canonical GR and it lies at the heart of general covariance, the principle on which GR is built.
The early results of \cite{hojman1976geometrodynamics}, where the vacuum was considered with the spatial metric being the only configuration variable, state that the Hamiltonian constraint, at second derivative order, is uniquely determined by the hypersurface deformation algebra (\ref{eq:Hypersurface deformation algebra - HaHa})-(\ref{eq:Hypersurface deformation algebra - HH}), and given by that of GR, up to the choice of Newton's and the cosmological constants \cite{hojman1976geometrodynamics,kuchar1974geometrodynamics}, implying that no modifications are allowed.
A crucial loophole to this conclusion lies in the common assumption that the spatial metric $q_{a b}$ must be a configuration variable.
Physically, this is the assumption that the metric is gravity itself, that it is a fundamental field.
This is the lesson we learned from GR: spacetime is a field itself, and it is the gravitational field.
But the recent detailed study of \cite{Covariance_regained} shows that this assumption is not necessary to obtain a field theory describing spacetime.
A fully consistent spacetime theory can be obtained by instead assuming a set of fundamental fields composing the phase-space, and considering non-classical constraints that still respect the form of the hypersurface deformation algebra (\ref{eq:Hypersurface deformation algebra - HaHa})-(\ref{eq:Hypersurface deformation algebra - HH}) up to the structure function $q^{a b}$ differing from the classical one.
This new structure function, composed by the fundamental variables of the phase space but not identical to any one of them, is interpreted as the inverse of the spatial metric.
This is an emergent spatial metric, and when embedded into a 4-dimensional manifold it gives rise to an emergent spacetime that is not gravity, but made of gravity.
This is the theory of emergent modified gravity (EMG).

We learned two key lessons on the nature of spacetime in \cite{Covariance_regained}: 1) the hypersurface deformation algebra can be used as a mechanism to obtain the spacetime metric in terms of the truly fundamental fields, 2) the anomaly-freedom of the constraint algebra does not imply general covariance.
These two lessons gave birth to EMG.
Lesson 1 tells us that spacetime is not gravity, but rather it is made of gravity, which can be used to weaken the assumptions that lead to the uniqueness results of \cite{hojman1976geometrodynamics}, and allows for modified gravity theories different from GR; we refer to the resulting spacetime metric as the \emph{emergent spacetime}.
Lesson 2 tells us that not all modifications allowed by the anomaly-freedom of the constraint algebra are indeed covariant, rather further \emph{covariance conditions} must be demanded to obtain a covariant modified gravity theory.

An explicit realization of EMG theories have so far been obtained only in the spherically symmetric reduced model \cite{Covariance_regained}.
An earlier, special case of EMG in spherical symmetry was studied in \cite{alonso2021anomaly,alonso2022nonsingular} where holonomy corrections were introduced motivated by loop quantum gravity (LQG) \cite{rovelli2004quantum,thiemann2008modern,bojowald2000symmetry} and shown to have a non-singular black hole solution.
The global structure of such a solution is an interuniversal wormhole joining a black hole to a white hole through their interiors.
Another application of spherically symmetric EMG is the covariant realization of modified Newtonian dynamics (MOND) as a solution to the dark matter puzzle \cite{milgrom1983modification,banik2022galactic,MONDEMG}.
Earlier attempts to obtain modified gravity theories in spherical symmetry include \cite{Tibrewala_Midisuperspace,Bojowald_DeformedGR,Tibrewala_Inhomogeneities} modelling inverse-triad corrections and holonomy corrections both motivated by LQG; however, only the former corrections are special cases of the EMG Hamiltonian, while the holonomy corrections of these last three references were found not to be covariant in \cite{Covariance_regained}.

EMG differs from other theories of emergent gravity in that the former does not consider the gravitational field as emergent.
Instead, in EMG the gravitational field is indeed fundamental, it is the spacetime that is emergent, such that the degrees of freedom of GR may be preserved in EMG.
One of the better known emergent gravity theories is entropic gravity \cite{EntropicGravity2011}\textemdash which is in turn inspired by the holographic principle and the AdS/CFT correspondence \cite{HolPrinc1,HolPrinc2,AdSCFT}\textemdash, where the gravitational force is a consequence of the temperature and change of entropy due to the change in the amount of information associated with the displacement of matter.
A similar stance is taken in \cite{Jacobson1995} which focuses on the thermodynamic nature of GR and argues that the Einstein equations are in some sense an equation of state derived from the proportionality of entropy and horizon area together with the fundamental thermodynamic relation $\delta Q= T {\rm d} S$.
It is, however, not clear what the degrees of freedom in entropic gravity and related theories really are, for instance, in the vacuum.

The explicit applications of EMG have so far been all in the vacuum, that is, where gravity is the only fundamental field.
Matter coupling in EMG is the extension of the vacuum theory to that of a larger phase space including matter fields.
Just as covariance conditions had to be placed on the emergent spacetime metric, covariance conditions have to be identified for the physical, possibly emergent, manifestations of the introduced matter fields.
Such detailed study would allow us to ensure that the matter coupling and possible modifications are indeed covariant.
This paper aims to address some of these subtleties on matter coupling in EMG, and it will be focused on the perfect fluid.

Despite being the 'simplest' form of a matter field, the perfect fluid couples in a simple, intuitive way only in the conventional, geometric approach to general relativity through the stress-energy tensor, but its extension to a Hamiltonian, or even Lagrangian formalism quickly becomes difficult.
Dust was introduced into canonical general relativity in \cite{brown1995dust} as timelike dust and in \cite{PhysRevD.56.4878} as null dust; the inclusion of pressure for a timelike perfect fluid was introduced in \cite{brown1993action} starting from the Lagrangian formulation followed by the ADM decomposition to obtain its canonical counterpart.
Because EMG has been formulated only in its canonical form and no action principle is available for the regaining process of the emergent spacetime, the introduction of the perfect fluid in purely canonical terms with no reliance on Lagrangians is paramount.
We do this explicitly in the present paper for both the timelike and null fluids by retaining the picture of the fluid as a collection of particles.
In particular, unlike \cite{brown1993action} where pressure $P (n , s)$ enters as a function of the particle number density $n$ and entropy density $s$, our analysis here shows that pressure can arise, not as $P(n,s)$, but as a function of the phase-space momenta in a non-trivial way with no reliance on the introduction of entropy as an independent phase-space variable.
The perfect fluid we derive here is then an example of an isentropic fluid in this context.

The main application of spherically symmetric EMG is to black holes and to gravitational collapse.
In GR, once a star exceeds the Tolman–Oppenheimer–Volkoff (TOV) limit \cite{PhysRev.55.374} its pressure is insufficient to stop the collapse of matter and a black hole is formed inevitably leading to the formation of a spacetime singularity at its center \cite{oppenheimer1939continued,PhysRevLett.14.57} commonly interpreted as a breakdown of GR as a valid description of spacetime in the high curvature regions.
On the other hand, the astrophysical observations indicate much of the black holes' exterior seems well described by GR which predicts the formation of horizons and their well-known properties implying that a black hole can grow by absorbing the collapsing matter, but it cannot shrink because nothing comes out of it.

One clue for the black hole's fate came with the discovery of Hawking radiation \cite{hawking1974black}.
This radiation carries some of the black hole's energy away, and therefore it shrinks or evaporates until it exhausts its mass.
Black hole evaporation then leads to the information loss paradox because the Hawking radiation has a thermal spectrum and does not carry any additional information about the matter that formed the black hole: the information is lost, violating unitarity of quantum mechanics \cite{kerr1963gravitational,israel1968event,hawking1975particle}.
Thus, pure evaporation of the black hole via Hawking radiation cannot be the full story unless we accept the information loss.

This paradox seems to be a pathological consequence of the assumptions for black hole evaporation: 1) Using classical gravity (specifically, GR), 2) neglecting back-reaction, 3) the Hawking-radiated matter is different from the collapsing matter \cite{Ashtekar_2005}.
A way out of the information loss paradox is thus to consider a deviation from these assumptions, for instance, by using a non-classical gravity theory or introducing quantum gravity effects.
Furthermore, quantum gravity effects are expected to play a significant role in the resolution of the classical singularities of black holes.
Under the assumption that these divergences are indeed resolved, new paradigms to the black hole's fate can be provided.
We will use EMG as the underlying theory replacing GR.
Furthermore, as will be explained in more detail later, there is a specific parameter in the general EMG spherically symmetric Hamiltonian that can be interpreted as effective elements of quantum gravity, in particular, to the holonomy variables used in LQG.
While such parameter belongs to a more general result of EMG, and we use LQG only as an interpretational tool, we may refer to it as a \textit{holonomy} or \textit{quantum} parameter.
The effects of this parameter are the ones responsible for the non-singular behaviour of the black hole solution in \cite{alonso2022nonsingular} and will also play a central role the new dynamics of the dust collapse solution we present here.
Also note that the general conclusions of singularity formation \cite{PhysRevLett.14.57} can potentially be circumvented in EMG because some assumptions of GR, including the positive energy theorem \cite{gibbons1983positive}, need not apply to the equations of EMG since these are different from Einstein's equations.

In the present paper, we obtain an exact solution to the collapse of dust in spherically symmetric EMG with non-trivial 'holonomy' parameter with the expectation that it will reveal new important properties to the above puzzles.
The resulting scenario is that of the matter falling in and producing a black hole.
The radius of the star then reaches a minimum value and 'bounces' back, emerging out  with the properties of a white hole.
To complete the global structure of the spacetime, this solution must be continuously glued to the exterior region, which must be a solution to the vacuum equations, by the shared boundary given by the star's radius.
The gluing process does not follow uniquely from solving the equations of motion and different consistent gluings can lead to different physical phenomena.
The two proposals for the outcome of the collapse we will focus on is the formation of wormholes rather than simple black holes as the result of the collapse \cite{PhysRev.48.73,PhysRevLett.61.1446}, and the transition from a collapsing black hole to an exploding white hole as a result of quantum gravity effects \cite{Ashtekar_2005,PhysRevD.92.104020}.

Within GR, traversable wormhole solutions can be obtained only by including matter with exotic properties such as negative energy density, but they are possible without exotic matter when deviations from GR are considered.
This is the case of modified GR with an $R^2$ action \cite{PhysRevD.92.043516}, as well as the example given above of spherically symmetric EMG \cite{alonso2022nonsingular}.
Furthermore, both examples show a wormhole solution in the vacuum.

The black-to-white-hole-transition proposal we will focus on is based on the work in \cite{Ashtekar_2005,PhysRevD.92.104020}, which can be described in a semiclassical treatment as follows.
At the start of the collapse of matter, we may trust the classical theory, and the black hole is formed provided the critical mass is exceeded.
At semiclassical regions characterized by higher curvature, the motion of matter and the gravitational field is modified due to quantum effects, possibly slowing down the collapse.
At the highest curvature regions, quantum effects are strong and the semiclassical treatment might break down so that full quantum gravity may be needed for a detailed result.
However, under the assumption that no physical divergences occur at the maximum curvature regions, the collapsing matter will cross the would-be classical singularity surface and bounce, continuing its journey outwards.
The matter that has crossed the would-be singularity is no longer collapsing, but expanding away from the black hole's center: this region then behaves as a white hole.
The matter will then exit the horizon and the black hole shrinks as it does so until the horizon disappears.
In the full quantum gravity context, this process can be understood as a quantum transition from a black hole to a white hole with a transition amplitude associated to it that will depend on the quantum gravity model used.
For an effort to compute this transition amplitude in the LQG approach, see \cite{Christodoulou_2016,Soltani_2021,D_Ambrosio_2021}.
Here we will focus on the semi-classical treatment using spherically symmetric EMG as the theory providing the effective quantum gravity equations.
Notice that this paradigm does not even require Hawking radiation, and the only major effects of its introduction would be to speed up the transition, and turn the collapsing and explosion phases asymmetric.

Previous results compatible with this paradigm include spherical models of null shells coupled to gravity where the classical solution collapsing into a black hole is connected to the (time-reversed) expanding solution emerging from a white hole via quantization \cite{hajivcek2001singularity} (see \cite{louko1998hamiltonian} for a canonical treatment of spherical null shells).
The classical metric describing the exterior of such null shell was explicitly obtained in \cite{PhysRevD.92.104020}.
The more recent work \cite{han2023geometry} studies the black-to-white-hole transition with the interior of the star described by an Oppenheimer-Snyder model modified by loop-quantum-cosmology (LQC) techniques \cite{bojowald2008loop}.
While such a model can be useful to give us insights on what LQG may predict about this process, one cannot rely on the LQC equations being covariant, since homogeneity makes it impossible to address such a question.
In this paper, we use the equations of EMG instead which are covariant by construction, and several technical features of the resulting process differ significantly from those of \cite{han2023geometry}.

In the last two examples, information falling into the black hole is not lost because in the wormhole proposal it would simply emerge out in the next universe, while in the black-to-white transition proposal it would emerge out of the white hole after the transition.
These two examples are then resolutions of the information loss paradox.

In this paper, we attempt to address all of the above issues in the context of EMG coupled to the perfect fluid.
The organization of this paper is as follows.
In Section~\ref{sec:Covariance in modified gravity} we review canonical gravity and EMG, and how the covariance conditions play a central role on defining the latter.
In Section~\ref{sec:Classical dust in canonical gravity} we briefly review how dust enters canonical gravity and identify the symmetries of the system associated to the dust that we will require to be preserved in EMG as additional conditions.
In Section~\ref{sec:The perfect fluid coupling} we proceed to couple the perfect fluid to EMG by applying the covariance conditions in canonical form.
We then obtain the explicit expression for the spherically symmetric Hamiltonian constraint for EMG in Section~\ref{sec:Spherically symmetric sector}.
In Section~\ref{eq:Gravitational collapse of timelike dust} we formulate the Oppenheimer-Snyder model in canonical terms and then focus on the gravitational collapse of dust in EMG, obtaining an exact solution.
Finally, in Section~\ref{sec:On the possible physical outcomes of the collapse} we discuss the possible physical outcomes of the collapse compatible with this solution.
We summarize the conclusions of this work in Section~\ref{sec:Conclusion}.

\section{Emergent modified gravity}
\label{sec:Covariance in modified gravity}

As usual in canonical theories, we assume that the spacetime, or the region of interest, is globally hyperbolic: $M = \Sigma \times \mathbb{R}$ with a 3-dimensional "spatial" manifold $\Sigma$.
Different choices of the embedding of $\Sigma$ in $M$ are parameterized by working with foliations of $M$ into smooth families of spacelike hypersurfaces $\Sigma_t$, $t\in{\mathbb R}$, each of which is homeomorphic to $\Sigma$.
For a given foliation, $\Sigma$ can be
embedded in $M$ as a constant-time hypersurface: $\Sigma\cong \Sigma_{t_0}\cong (\Sigma_{t_0},t_0)\hookrightarrow M$ for any fixed $t_0$.

Given a foliation into spacelike hypersurfaces $\Sigma_t$, a metric
$g_{\mu\nu}$ on $M$ defines the unit normal vector field $n^{\mu}$ on $\Sigma_{t_0}$, and induces a unique spacelike metric $q_{ab}(t_0)$ on
$\Sigma_{t_0}$ by restricting the space-time tensor $q_{\mu\nu}=g_{\mu\nu}+n_{\mu}n_{\nu}$ to vector fields tangential to $\Sigma_{t_0}$, while $q_{\mu\nu}n^{\nu}=0$. (We use Greek letters for indices of spacetime tensors, and Latin letters for indices of spatial tensors.) 
Time-evolution then provides a family of spatial metrics, one for each hypersurface.
Unambiguous evolution requires an additional structure that relates points between infinitesimally adjacent hypersurfaces, and this is provided by a time-evolution vector field
\begin{equation}
  t^\mu = N n^\mu + N^a s_a^\mu 
  \label{eq:Time-evolution vector field}
\end{equation}
in spacetime, with the lapse function $N$ and shift vector field $N^a$
\cite{arnowitt2008republication}.
The three vector fields $s_a^\mu(t_0)$ inject $T\Sigma_{t_0}$ into $TM$ such that $g_{\mu \nu} n^\mu s^\mu_a = 0$, playing the role of the spatial basis vectors on the spatial hypersurfaces.
The lapse and shift describe, via (\ref{eq:Time-evolution vector field}), the frame of an observer in curved spacetime who measures the physical fields evolving on the hypersurfaces.
The resulting spacetime metric or line element is then given by \cite{arnowitt2008republication}
\begin{equation}
  {\rm d} s^2 = - N^2 {\rm d} t^2 + q_{a b} ( {\rm d} x^a + N^a {\rm d} t )
  ( {\rm d} x^b + N^b {\rm d} t )
  \ .
  \label{eq:ADM line element}
\end{equation}

Time-evolution and gauge transformations are described by the same flow via Poisson brackets, generated by the Hamiltonian constraint $H$ and the diffeomorphism constraint $H_a$ differing only by their smearing function.
Our notation here denotes $H$ and $H_a$ as the local versions of the constraints and we use the square brackets to denote smearing by the function in the argument, i.e., $H[N]=\int {\rm d} x^3 H(x) N(x)$.
Given a phase-space function $\mathcal{O}$, its infinitesimal gauge transformation is given by $\delta_\epsilon \mathcal{O} = \{ \mathcal{O} , H[\epsilon^0 , \epsilon^a]
\}$, where $H[\epsilon^0 , \epsilon^a] = H[\epsilon^0] + H_a [\epsilon^a]$, and $\epsilon^0$ and $\epsilon^a$ are infinitesimal gauge parameters.
The infinitesimal time-evolution, on the other hand, is given by $\Dot{\mathcal{O}} = \delta_t \mathcal{O} = \{ \mathcal{O} , H[N , N^a] \}$, that is, lapse and shift play the role of gauge parameters in the time-flow.
In its role as the gauge flow generator, $H[\epsilon^0 , \epsilon^a]$ must be a constraint $H[\epsilon^0 , \epsilon^a]=0$ for all $\epsilon^0$ and $\epsilon^a$.
Therefore, the dynamics is constrained too: $H[N , N^a]=0$.
We say that the physical solutions of the theory are "on-shell," that is, the phase-space variables on each hypersurface that solve the equations of motion are such that the constraints vanish, $H[N]=0$ and $H_a[N^a]=0$.

The poisson brackets do not directly provide the gauge transformations of $N$ and $N^a$ because they do not have momenta, physically this means that they do not evolve dynamically but rather they specify the frame with respect to which evolution is defined.
Instead, the gauge transformations of the lapse and shift are derived from the condition that the form of the equations of motion of the phase-space are gauge invariant.
The gauge transformations obeying this condition are given by \cite{pons1997gauge,salisbury1983realization,bojowald2018effective}
\begin{eqnarray}
    \delta_\epsilon N &=& \dot{\epsilon}^0 + \epsilon^a \partial_a N - N^a \partial_a \epsilon^0
    \ ,
    \label{eq:Off-shell gauge transformations for lapse}
    \\
    \delta_\epsilon N^a &=& \dot{\epsilon}^a + \epsilon^b \partial_b N^a - N^b \partial_b \epsilon^a + q^{a b} \left(\epsilon^0 \partial_b N - N \partial_b \epsilon^0 \right)
    \ .
    \label{eq:Off-shell gauge transformations for shift}
\end{eqnarray}

We say that the spacetime (\ref{eq:ADM line element}) is covariant if and only if
\begin{equation}
    \delta_\epsilon g_{\mu \nu} \big|_{\text{O.S.}} =
    \mathcal{L}_{\xi} g_{\mu \nu} \big|_{\text{O.S.}}
    \ ,
    \label{eq:Covariance condition of spacetime}
\end{equation}
where "O.S." means we evaluate the expression on-shell.
The content of (\ref{eq:Covariance condition of spacetime}) is the condition that the canonical gauge transformations with the gauge parameters $(\epsilon^0, \epsilon^a)$ reproduce diffeomorphisms of the spacetime metric with the 4-vector generator $\xi^\mu$ related by
\begin{eqnarray}
    \xi^\mu &=& \epsilon^0 n^\mu + \epsilon^a s^\mu_a
    = \xi^t t^\mu + \xi^a s^\mu_a
    \ ,
    \\
    \xi^t &=& \frac{\epsilon^0}{N}
    \ ,
    \hspace{0.75cm}
    \xi^a = \epsilon^a - \frac{\epsilon^0}{N} N^a
    \ ,
\label{eq:Diffeomorphism generator projection}
\end{eqnarray}

The timelike components ($tt$ and $ta$) of the spacetime covariance condition (\ref{eq:Covariance condition of spacetime}) are automatically satisfied by the gauge transformations of the lapse and shift, (\ref{eq:Off-shell gauge transformations for lapse}) and (\ref{eq:Off-shell gauge transformations for shift}), provided the covariance condition of the 3-metric, $\delta_\epsilon q_{a b} |_{\text{O.S.}} = \mathcal{L}_{\xi} q_{a b} |_{\text{O.S.}}$, is satisfied too.
The latter, contrary to what is commonly stated, is not automatically satisfied just by virtue of the hypersurface deformation algebra; it can be simplified to the following series of conditions \cite{Covariance_regained}
\begin{equation}
    \frac{\partial (\delta_{\epsilon^0} q^{a b})}{\partial (\partial_c \epsilon^0)} \bigg|_{\text{O.S.}}
    = \frac{\partial (\delta_{\epsilon^0} q^{a b})}{\partial (\partial_c \partial_d \epsilon^0)} \bigg|_{\text{O.S.}}
    = \dotsi
    = 0
    \ ,
    \label{eq:Covariance condition of 3-metric - reduced}
\end{equation}
where the series terminates on the highest derivative order considered in the Hamiltonian constraint, which here we assume to be finite\textemdash that is, we assume a local theory.

If the phase-space is composed of the spatial metric $q_{a b}$ (the inverse of the structure function) as a configuration variable and its conjugate momenta $p^{a b}$, then the covariance condition (\ref{eq:Covariance condition of 3-metric - reduced}) implies, from
$\{q_{a b} , H[\epsilon^0]\} = \delta H[\epsilon^0] / \delta p^{a b}$, that
the Hamiltonian constraint must not contain spatial derivatives of $p^{a b}$.
If we use only up to second-order spatial derivatives of $q_{ab}$, the Hamiltonian constraint is uniquely determined by the hypersurface deformation algebra (\ref{eq:Hypersurface deformation algebra - HaHa})-(\ref{eq:Hypersurface deformation algebra - HH}) up to the choice of Newton's and the cosmological
constant \cite{hojman1976geometrodynamics,kuchar1974geometrodynamics}.
It must therefore be the classical constraint of GR, and generally covariant modifications are ruled out under the stated conditions.

Here is where EMG differs from traditional canonical gravity.
The assumption that the spatial metric $q_{ab}$ is a configuration variable of the phase-space is not necessary to obtain a field theory describing spacetime, and we may drop it altogether, except when trying to recover GR.
Instead, we assume that the phase-space is composed of certain fundamental fields different from the metric, and the metric is an emergent object to be regained by the following process.
Leaving the diffeomorphism constraint unmodified, we allow for the Hamiltonian constraint to deviate from its classical expression and we say it is modified.
In vacuum, the only fundamental field is gravity.
The Hamiltonian constraint in vacuum is then restricted to satisfy a hypersurface deformation algebra (\ref{eq:Hypersurface deformation algebra - HaHa})-(\ref{eq:Hypersurface deformation algebra - HH}) where the structure function is allowed to be different (and not just related by a simple canonical transformation) from its classical expression.
The inverse of the structure function obtained from such a procedure now plays the role of the \emph{emergent} spatial metric and it is used in (\ref{eq:ADM line element}) to define the \emph{emergent} spacetime metric.
The last step is to demand that this emergent metric and the modified constraint satisfy the covariance condition (\ref{eq:Covariance condition of 3-metric - reduced}).
This is EMG: a covariant theory of an emergent spacetime.

If one considers additional matter fields into the theory that present manifestations independently from the spacetime metric, then one has to make sure that such manifestations of the matter fields are covariant too.
For example, if the matter field in consideration is described by some spacetime tensor $f$, then one has to apply the matter covariance condition on this field too:
\begin{equation}
    \delta_\epsilon f |_{\text{O.S.}} = \mathcal{L}_\xi f |_{\text{O.S.}}
    \ .
\end{equation}
Because the spacetime metric is emergent, then it is allowed not only to depend on gravity, but also on the matter fields as long as the anomaly-freedom of the constraint algebra and all the covariance conditions are satisfied.
Here we will be interested in the perfect fluid by envisioning it as a collection of particles in the (emergent) spacetime, thus, it is fundamentally described by its 4-velocity $u^\mu$ (or co-velocity $u_\mu$).
The matter covariance condition must then be applied to this quantity.
For completeness, we will also place the covariance condition on the energy density such that it transforms as a spacetime scalar.

\section{Classical dust in canonical gravity}
\label{sec:Classical dust in canonical gravity}

\subsection{Pressureless dust}

Unlike other matter fields, it is challenging to treat the perfect fluid\textemdash a model for collective particles\textemdash in both the Lagrangian and canonical formulations of general relativity due to the ambiguities related to the equation of state.
This ambiguity stems from the pressure function not having a clear dependence on the generalized coordinates or the phase-space variables of the Lagrangian and canonical formulations, respectively. In \cite{brown1993action}, for instance, the pressure $P (n , s)$ is postulated as a function of the particle number density $n$ and a new phase-space variable $s$ they called entropy density. Our analysis here shows that pressure can be derived rather than postulated without introducing the entropy density $s$ and is therefore an example of an isentropic perfect fluid in this context. Furthermore, as we will see below, the pressure function we derived here does not depend solely on the number density $n$, but rather on the 'ratio' between the fluid's radial and time momenta as will become clear in Section~\ref{sec:The perfect fluid coupling}.
If one neglects the pressure, however, the fluid is called dust and it is relatively easy to treat in both formulations, so this will be our starting point.

In the canonical formulation, the dust field is described by the coordinate fields $T(x)$ and $Z^i(x)$ (with $i= 1,2,3$) of its collective particles of rest mass $\mu$ as the configuration variables\textemdash note that these are different from the coordinates $x$ of the manifold\textemdash, and their respective conjugate momenta are denoted by $P^{(T)} (x)$ and $P^{(Z)}_i (x)$, representing the usual (densitized) energy and linear momentum (density) of the particles.
The resulting Poisson brackets are therefore
\begin{eqnarray}
    \{T (\Vec{x}) , P^{(T)} (\Vec{y})\} &=& \delta^{3} (\Vec{x} - \Vec{y})
    \ , \nonumber\\
    \{Z^i (\Vec{x}) , P^{(Z)}_j (\Vec{y})\} &=& \delta^i_j \delta^{3} (\Vec{x} - \Vec{y})
    \ .
\end{eqnarray}
When coupled to gravity, the four canonical pairs $(T,P^{(T)})$, $(Z^i,P^{(Z)}_i)$ add four degrees of freedom to the theory.

The configuration variables define a 1-form ${\rm d} u$, describing the co-velocity field of the dust particles with components $u_\mu \equiv - \partial_\mu T - W_i \partial_\mu Z^i$, where $W_i = P^{(Z)}_i / P^{(T)}$ is the (internal) 3-velocity.
The co-velocity field satisfies the normalization $g^{\mu \nu} u_\mu u_\nu = - s$, where one picks $s=1$ for timelike dust and $s=0$ for null dust.

The classical diffeomorphism and Hamiltonian constraints contributions of the dust respectively are \cite{brown1995dust,PhysRevD.56.4878}
\begin{eqnarray}
    H^{\text{matter}}_a &=&
    P^{(T)} \partial_a T
    + P^{(Z)}_i \partial_a Z^i
    \ ,
    \label{eq:Dust contributions to diffeomorphism constraint}
    \\
    H_s^{\rm dust} &=&
    \sqrt{s (P^{(T)})^2 + q^{a b} H^{\rm matter}_a H^{\rm matter}_b}
    \ ,
    \label{eq:Dust contributions to Hamiltonian constraint}
\end{eqnarray}
where $q^{ab}$ is the structure function of the constraint algebra (\ref{eq:Hypersurface deformation algebra - HH}).
Energy quantities as observed in the Eulerian frame adapted to the foliation (observers with 4-velocity $n^\mu$) can be obtained directly from the constraints.
The Eulerian energy density $\rho_s^{\rm (E)}$ is given by
\begin{equation}
    \sqrt{\det q} \rho_s^{\rm (E)} \equiv \frac{\delta H_s^{\rm dust} [N]}{\delta N}
    = H^{\rm dust}_s
    = P^{(T)} \sqrt{s + q^{ab} u_a u_b}
    \ ,
\end{equation}
and the Eulerian dust current $J_a^{\rm (E)}$ by
\begin{equation}
    \sqrt{\det q} J_a^{\rm (E)} \equiv - \frac{\delta \vec{H}^{\rm matter} [\vec{N}]}{\delta N^a}
    = - H^{\rm matter}_a
    = P^{(T)} u_a
    \ ,
\end{equation}
which can be extended to the 4-current $\sqrt{\det q} J_\mu^{\rm (E)} = P^{(T)} u_\mu$, where the normal component yields the mass density of the dust.
Using the particle rest mass $\mu$, we can then obtain the Eulerian particle number density,
\begin{equation}
    \mu n^{\rm (E)}_s \equiv - n^\nu J_\nu
    = \frac{P^{(T)}}{\sqrt{\det q}}
    \ .
    \label{eq:Eulerian particle number density - dust}
\end{equation}
This result confirms our interpretation of $P^{(T)}$ as the (densitized) mass density.
The Eulerian spatial stress tensor is given by
\begin{equation}
    \sqrt{\det q} S_{a b}^{\rm (E)} \equiv \frac{2}{N} \frac{\delta H_s^{\rm dust} [N]}{\delta q^{a b}}
    = \frac{(P^{(T)})^2}{H_s^{\rm dust}} u_a u_b
    \ ,
\end{equation}
and the Eulerian pressure by
\begin{equation}
    P^{\rm (E)} \equiv \frac{q^{a b} S_{a b}}{3}
    = \frac{1}{3} \frac{(P^{(T)})^2}{H_s^{\rm dust}} q^{a b} u_a u_b
    \ .
\end{equation}

The dust has a relative velocity with respect to Eulerian observers such that it is boosted with respect to the them by the Lorentz factor
\begin{eqnarray}
    \gamma^{\rm (E)} &=& - n^\mu u_\mu
    = \frac{\sqrt{s (P^{(T)})^2 + q^{a b} H^{\rm matter}_a H^{\rm matter}_b}}{P^{(T)}}
    \nonumber\\
    &=& \sqrt{s + q^{a b} u_a u_b}
    \ ,
\end{eqnarray}
where we used the equations of motion for $\dot{T}$ and $\dot{Z}^i$.
We can use this boost factor to obtain the energy density in the dust frame,
\begin{equation}
    \rho_s^{\rm dust} = (\gamma^{\rm (E)})^{-2} \rho^{\rm (E)}_s
    = \frac{1}{\sqrt{\det q}} \frac{(P^{(T)})^2}{H^{\rm dust}_s}
    \ .
    \label{eq:Dust energy density}
\end{equation}

\subsection{Classical symmetries}

The dust constraints (\ref{eq:Dust contributions to diffeomorphism constraint})-(\ref{eq:Dust contributions to Hamiltonian constraint}) have the following important symmetry generators.

The phase-space function
\begin{equation}
    Q_0 [\alpha] = \int {\rm d}^3 x\ \alpha P^{(T)}
    = \int {\rm d}^3 x\ \alpha \sqrt{\det q} \mu n^{\rm (E)}_s
    \ ,
    \label{eq:Perfect fluid symmetry generator - total mass}
\end{equation}
where $\alpha$ is a constant, and we used (\ref{eq:Eulerian particle number density - dust}) to write the second equality, commutes with the dust constraints $\{Q_0[\alpha],H^{\rm dust}[N]\}=\{Q_0[\alpha],\vec{H}^{\rm matter}[\vec{N}]\}=0$.
This implies that $\dot{Q}_0[\alpha]=0$, which in turn implies that (\ref{eq:Perfect fluid symmetry generator - total mass}) is a conserved global charge.
Taking $\alpha=1$, we identify this charge as the total mass of the collective particles in the Eulerian frame, implying conservation of Eulerian mass and particle number.

A second symmetry generator of the dust constraints is given by the three global charges
\begin{equation}
    Q_i [\beta^i] = \int {\rm d}^3 x\ \beta^i P^{(Z)}_i
    = \int {\rm d}^3 x\ \sqrt{\det q} \beta^i W_i \mu n^{\rm (E)}_s
    \ ,
    \label{eq:Perfect fluid symmetry generator - total mass flow}
\end{equation}
where $\beta^i = (\beta^1 , \beta^2 , \beta^3)$ are constants.
Choosing a unit internal vector $\beta^j = \hat{\beta}^j$, we identify this quantity as the total linear mass-flux component in the direction $\hat{\beta}^j$.

The third symmetry generator of the dust constraints is related to its SO(3) global symmetry corresponding to the internal rotation of the dust variables.
In particular, this infinitesimal transformation takes the form
\begin{eqnarray}
    Z^i &\to&
    Z^i + \tensor{\epsilon}{^i_j_k} \theta^j Z^k
    \ , \\
    P^{(Z)}_i &\to&
    P^{(Z)}_i + \tensor{\epsilon}{_i_j^k} \theta^j P^{(Z)}_k
    \ ,
\end{eqnarray}
where $\theta^i = (\theta^1 , \theta^2 , \theta^3)$ are constant parameters denoting the angle of rotation along an internal axis with direction $\hat{\theta}^i$, the totally antisymmetric tensor $\epsilon_{i j k}$ (where $\epsilon_{123}=1$) is the Lie algebra generator $\tau_i \in \mathfrak{so}(3)$ in the defining representation, and we rise (and lower) internal indices with the Kronecker delta $\delta^{ij}$ (and $\delta_{ij}$).
This transformation is generated by the phase-space function
\begin{eqnarray}
    G_j [\theta^j] &=& \int {\rm d}^3 x \ \theta^j \tensor{\epsilon}{_j_k^i} Z^k P^{(Z)}_i
    \nonumber\\
    &=& \int {\rm d}^3 x\ \sqrt{\det q} \left( \theta^j \tensor{\epsilon}{_j_k^i} Z^k \mu W_i n^{\rm (E)}_s \right)
    \nonumber\\
    &=& \int {\rm d}^3 x\ \sqrt{\det q} \left( \vec{\theta} \cdot \vec{Z} \times (\vec{W} \mu n^{\rm (E)}_s) \right)
    \ .
    \label{eq:Perfect fluid symmetry generator - SO(3)}
\end{eqnarray}
Choosing a unit internal vector $\theta^j = \hat{\theta}^j$, we identify this quantity as the total angular mass-flux component in the direction $\hat{\theta}^j$.

While the dust constraints (\ref{eq:Dust contributions to diffeomorphism constraint})-(\ref{eq:Dust contributions to Hamiltonian constraint}) contain even more symmetries, the three global charges (\ref{eq:Perfect fluid symmetry generator - total mass}), (\ref{eq:Perfect fluid symmetry generator - total mass flow}), and (\ref{eq:Perfect fluid symmetry generator - SO(3)}) are basic conserved quantities that must hold in the generalization to the perfect fluid (at least for pressure functions that arise from conservative interactions between the fluid particles) and, furthermore, in EMG too since we want to retain the picture of the perfect fluid as a collection of particles.

In the following  we will re-derive the dust Hamiltonian constraint $H_s^{\text{dust}}$ from its basic properties such as normalization and covariance conditions, and we will then generalize it to that of the perfect fluid which includes the pressure in the form of an equation of state even in the context of EMG.
We will refer to this more general constraint contribution as $H^{\text{matter}}_s$.
The steps of this procedure are the following.
We demand that the full (modified) constraints $H = H^{\text{grav}} + H^{\text{matter}}_s$ and $H_a = H^{\text{grav}}_a + H^{\text{matter}}_a$ form an anomaly-free hypersurface deformation algebra with (emergent) structure function $q^{a b}$; 
we demand that the (emergent) spacetime metric $g_{\mu \nu}$ is covariant, as well as the co-velocity of the fluid $u_\mu$ for which a further covariance condition will be formulated in the next section; we demand that the fluid's velocity is normalized; and, finally, we demand that the phase space functions (\ref{eq:Perfect fluid symmetry generator - total mass}), (\ref{eq:Perfect fluid symmetry generator - total mass flow}), and (\ref{eq:Perfect fluid symmetry generator - SO(3)}) remain as symmetry generators in the modified theory, thus preserving the conserved quantities and their physical meaning.
It is easier to consistently realize all of these demands in the reverse order due to their increasing difficulty, and this is the approach we will take in the following sections.
After imposing these conditions, we will solve them exactly starting with the most general constraint ansatz with arbitrary dependence on the fluid's phase-space variables and the first-order derivatives of its configuration variables, while also allowing the structure function and the gravitational constraint contribution to depend on the fluid's configuration variables.
Lastly, we will place a covariance condition on the energy density $\rho_s$ such that it transforms as a spacetime scalar.
We will also show that the pressure $P$ will appear as an emergent property of the fluid and its covariant transformation as a scalar will be implied by those of $g_{\mu \nu}$, $u_\mu$, and $\rho_s$.
We will do all of this in purely canonical grounds so that an underlying action is not necessary for the existence of the covariant Hamiltonian and, furthermore, so that the results hold for EMG too.

\section{The perfect fluid coupling}
\label{sec:The perfect fluid coupling}

Throughout the first half of this section we assume that the spacetime covariance condition is already satisfied because it simplifies the analysis of the other conditions. There is no loss of generality in making this assumption because one must still implement it when deriving the explicit Hamiltonian constraint. We return to this condition in the Subsection~\ref{sec:Anomaly-freedom and spacetime covariance}.

\subsection{Symmetry conditions}

We start by defining our Hamiltonian constraint ansatz in such a way that we can split the constraint in the form $H = H^{\rm grav} + H_s^{\rm matter}$ where they have the dependence
\begin{eqnarray}
    && H^{\rm grav} \left( T , Z^i \right)
    \ , \nonumber\\
    && H_s^{\rm matter} \left( T , Z^i , P^{(T)} , P^{(Z)}_i , \partial_a T , \partial_a Z^i \right)
    \ ,
    \label{eq:Hamiltonian constraint ansatz}
\end{eqnarray}
where we have suppressed the possible dependence on the gravitational variables to ease the notation.
We also assume that in the vacuum limit defined by $T, Z^i , P^{(T)} , P^{(Z)}_i \to 0$, we obtain $H \to H^{\rm grav}$ with the latter depending only on the gravitational variables\textemdash therefore, $H^{\rm grav}$ has a more complicated dependence on the gravitational variables than $H_s^{\rm matter}$ does.

We now demand that the constraint ansatz (\ref{eq:Hamiltonian constraint ansatz}) commutes with the symmetry generators (\ref{eq:Perfect fluid symmetry generator - total mass}), (\ref{eq:Perfect fluid symmetry generator - total mass flow}), and (\ref{eq:Perfect fluid symmetry generator - SO(3)}).
Due to the common complexity of $H^{\rm grav}$, and in view that it depends on the gravitational variables much more heavily than $H_s^{\rm matter}$ does, we will assume that the symmetry generators (\ref{eq:Perfect fluid symmetry generator - total mass}), (\ref{eq:Perfect fluid symmetry generator - total mass flow}), and (\ref{eq:Perfect fluid symmetry generator - SO(3)}) must commute with $H^{\rm grav}$ and $H_s^{\rm matter}$ independently.

We obtain
\begin{eqnarray}
    \{ Q_0 [\alpha] , H [\epsilon^0] \} &=&
    - \int {\rm d}^3 x\ \alpha \epsilon^0 \frac{\partial H}{\partial T}
    \ , \\
    \{ Q_i [\beta^i] , H [\epsilon^0] \} &=&
    - \int {\rm d}^3 x\ \beta^i \epsilon^0 \frac{\partial H}{\partial Z^i}
\end{eqnarray}
where we have integrated out the first-order derivative term and neglected boundary terms.
Thus, $\{ Q_0 [\alpha] , H^{\rm grav} [\epsilon^0] \} = 0$ and $\{ Q_i [\beta^i] , H_s^{\rm matter} [\epsilon^0] \}=0$, for arbitrary $\alpha$, $\beta^i$, and $\epsilon^0$, reduce the phase-space dependence of the constraint ansatz (\ref{eq:Hamiltonian constraint ansatz}) to
\begin{equation}
    H_s^{\rm matter} \left( P^{(T)} , P^{(Z)}_i , \partial_a T , \partial_a Z^i \right)
    \ ,
\end{equation}
with $H^{\rm grav}$ now completely independent of the fluid's variables.
Therefore, $H^{\rm grav}$ is identical to the vacuum Hamiltonian constraint.

Similarly, the implementation of the symmetry generator (\ref{eq:Perfect fluid symmetry generator - SO(3)}) further reduces the phase-space dependence of the constraint ansatz to
\begin{equation}
    H_s^{\rm matter} \left( P^{(T)} , \partial_a T , P^{(Z)}_i \partial_a Z^i , (\vec{P}^{(Z)})^2 , (\partial \vec{Z})^2 \right)
    \ ,
    \label{eq:Constraint ansatz - symmetry generator}
\end{equation}
where $(\vec{P}^{(Z)})^2=\delta^{i j} P^{(Z)}_i P^{(Z)}_j$ and $(\partial \vec{Z})^2=\delta_{i j} \partial_a Z^i \partial_b Z^j$.


\subsection{Normalization condition}

The ADM decomposition of the normalization $g^{\alpha \beta} u_\alpha u_\beta = - s$, where we use the emergent inverse metric for this expression, can be rewritten as
\begin{equation}
    u_t
    =
    N^b u_b
    - N \sqrt{s + q^{a b} u_a u_b}
    \ ,
    \label{eq:Normalization condition - Time component of velocity}
\end{equation}
where the sign of the square root was chosen to preserve a negative $u_t$ (that is, future-pointing) even in the case $N^b = 0$.
Here, $q^{a b}$ is the structure function of the resulting hypersurface deformation algebra, that is, the emergent one.

Using the expressions
\begin{eqnarray}
    u_t 
    &=&
    - \dot{T}
    - \frac{P^{(Z)}_i}{P^{(T)}} \dot{Z}^i
    \ , \\
    u_a
    &=&
    - H_a^{\text{matter}} / P^{(T)}
    \ ,
\end{eqnarray}
the normalization places a restriction on the equations of motion of the dust:
Taking Hamilton's equations of motion $\dot{T} = \{ T , H [N , \vec{N}] \}$\textemdash, and similarly for $Z^i$\textemdash with unmodified diffeomorphism constraint, the normalization expression (\ref{eq:Normalization condition - Time component of velocity}) can be written as
\begin{eqnarray}
    P^{(T)} \{ T , H [N] \}
    + P^{(Z)}_i \{ Z^i , H [N] \}
    &=&
    N \sqrt{s (P^{(T)})^2 + q^{a b} H^{\rm matter}_a H^{\rm matter}_b}
    \nonumber\\
    &=:&
    N H_s^{\rm dust}
    \ ,
    \label{eq:Normalization condition}
\end{eqnarray}
where $H^{\rm dust}_s$ differs from the classical expression (\ref{eq:Dust contributions to Hamiltonian constraint}) in that the classical structure function $q^{ab}$ may be replaced by the emergent one.
We shall use $H^{\rm dust}_s$ in this context in the following.

Using the constraint ansatz compatible with the symmetry conditions given by (\ref{eq:Constraint ansatz - symmetry generator}), the equation (\ref{eq:Normalization condition}), which must be satisfied for arbitrary $N$, simplifies into the condition
\begin{equation}
    P^{(T)} \frac{{\rm d} H}{{\rm d} P^{(T)}}
    + P^{(Z)}_i \frac{{\rm d} H}{{\rm d} P^{(Z)}_i}
    =
    H_s^{\rm dust}
    \ ,
    \label{eq:Normalization condition - zeroth term}
\end{equation}
where ${\rm d}$ stands for total derivative.
The solution to the normalization condition then reduces the phase-space dependence of the ansatz (\ref{eq:Constraint ansatz - symmetry generator}) to
\begin{equation}
    H_s^{\rm matter}
    =
    H_s^{\rm dust}
    - \bar{f} \left( (P^{(Z)})^2 / (P^{(T)})^2 , \partial_a T , (\partial \vec{Z})^2 \right)
    \ ,
    \label{eq:Constraint ansatz - symmetry, normalized}
\end{equation}
for some undetermined function $\bar{f}$.


\subsection{Matter covariance conditions I}

We now impose the covariance condition on the fluid's co-velocity field:
\begin{equation}
    \delta_\epsilon u_\mu \big|_{\text{O.S.}} = \mathcal{L}_\xi u_\mu \big|_{\text{O.S.}}
    \ .
    \label{eq:Covariance condition - Perfect fluid}
\end{equation}
To this end we perform the ADM decomposition of its infinitesimal coordinate transformation:
\begin{eqnarray}
    \mathcal{L}_\xi u_\mu
    &=& \frac{\epsilon^0}{N} \dot{u}_\mu
    + \left(\epsilon^a - \frac{\epsilon^0}{N} N^a\right) \partial_a u_\mu
    \nonumber\\
    &&
    + u_t \partial_\mu \left(\frac{\epsilon^0}{N}\right)
    + u_a \partial_\mu \left(\epsilon^a - \frac{\epsilon^0}{N} N^a\right)
    \ .
\end{eqnarray}
Explicitly, the components are
\begin{eqnarray}
    \mathcal{L}_\xi u_a
    &=&
    \frac{\epsilon^0}{N} \left( \dot{u}_a
    - \frac{u_t
    - N^b u_b}{N} \partial_a N
    - \left( N^b \partial_b u_a
    + u_b \partial_a N^b \right) \right)
    + \frac{u_t
    - N^b u_b}{N} \partial_a \epsilon^0
    + \epsilon^b \partial_b u_a
    + u_b \partial_a \epsilon^b
    \nonumber\\
    &=&
    \frac{\epsilon^0}{N} \left( \dot{u}_a
    + \sqrt{s + q^{a b} u_a u_b} \partial_a N
    - \left( N^b \partial_b u_a
    + u_b \partial_a N^b \right) \right)
    - \sqrt{s + q^{a b} u_a u_b} \partial_a \epsilon^0
    + \epsilon^b \partial_b u_a
    + u_b \partial_a \epsilon^b
    \ ,
    \label{eq:r.h.s covariance conditon - spatial component}
\end{eqnarray}
\begin{eqnarray}
    \mathcal{L}_\xi u_t
    &=& \frac{\epsilon^0}{N} \partial_t \left( u_t
    - u_a N^a \right)
    + \left(\epsilon^a - \frac{\epsilon^0}{N} N^a\right) \partial_a u_t
    + \frac{u_t - N^a u_a}{N} \left( \dot{\epsilon}^0
    - \frac{\epsilon^0}{N} \dot{N} \right)
    + u_a \dot{\epsilon}^a
    + \frac{\epsilon^0}{N} N^a \dot{u}_a
    \notag\\
    &=&
    - \epsilon^0 \partial_t \sqrt{s + q^{a b} u_a u_b}
    + \left(\epsilon^a - \frac{\epsilon^0}{N} N^a\right) \partial_a u_t
    - \sqrt{s + q^{a b} u_a u_b} \dot{\epsilon}^0
    + u_a \dot{\epsilon}^a
    \ ,
    \label{eq:r.h.s covariance conditon - time component}
\end{eqnarray}
where we used the normalization expression (\ref{eq:Normalization condition - Time component of velocity}) to simplify these components into their second lines.

Also, the following calculation will be useful:
\begin{eqnarray}
    \{ u_a , H [N]\}
    &=&
    - \partial_a \left( \{ T , H [N]\} \right)
    - \frac{P^{(Z)}_i}{P^{(T)}} \partial_a \left( \{ Z^i , H [N]\} \right)
    - \left\{ \frac{P^{(Z)}_i}{P^{(T)}} , H [N] \right\} \partial_a Z^i
    \notag\\
    &=& 
    - \partial_a \left( \{ T , H [N]\}
    + \frac{P^{(Z)}_i}{P^{(T)}} \{ Z^i , H [N]\} \right)
    + \{ Z^i , H [N]\} \partial_a \left( \frac{P^{(Z)}_i}{P^{(T)}} \right)
    - \left\{ \frac{P^{(Z)}_i}{P^{(T)}} , H [N] \right\} \partial_a Z^i
    \notag\\
    &=& 
    - \partial_a \left( \frac{N H_s^{\text{dust}}}{P^{(T)}} \right)
    + \{ Z^i , H [N]\} \partial_a \left( \frac{P^{(Z)}_i}{P^{(T)}} \right)
    - \left\{ \frac{P^{(Z)}_i}{P^{(T)}} , H [N] \right\} \partial_a Z^i
    \ ,
    \label{eq:4-velocity components - explicit gauge transformation}
\end{eqnarray}
where we used the normalization condition (\ref{eq:Normalization condition}) in the last line.

We will now focus on the spatial component of the covariance condition (\ref{eq:r.h.s covariance conditon - spatial component}).
Using Hamilton's equations of motion $\dot{u}_a = \{ u_a , H [N] + H_b [N^b] \}$, (\ref{eq:r.h.s covariance conditon - spatial component}) becomes
\begin{equation}
    \frac{1}{\epsilon^0} \left( \frac{P^{(T)}}{H^{\rm dust}_s} \{ u_a , H [\epsilon^0]\}
    + \partial_a \epsilon^0 \right) \bigg|_{\text{O.S.}}
    =
    \frac{1}{N} \left( \frac{P^{(T)}}{H^{\rm dust}_s} \{ u_a , H [N]\}
    + \partial_a N \right)
    \bigg|_{\text{O.S.}}
    \ .
    \label{eq:Covariance conditon in Poisson brackets - spatial component}
\end{equation}
Using \eqref{eq:4-velocity components - explicit gauge transformation} it can be rewritten into
\begin{equation}
    \frac{1}{\epsilon^0} \left(\{ Z^i , H [\epsilon^0]\} \partial_a W_i
    - \left\{ W_i , H [\epsilon^0] \right\} \partial_a Z^i\right) \big|_{\text{O.S.}}
    \nonumber\\
    =
    \frac{1}{N} \left(\{ Z^i , H [N]\} \partial_a W_i
    - \left\{ W_i , H [N] \right\} \partial_a Z^i\right)
    \big|_{\text{O.S.}}
    \ .
    \label{eq:Covariance conditon in Poisson brackets - spatial component - simplified}
\end{equation}
Upon substitution of the ansatz (\ref{eq:Constraint ansatz - symmetry generator}), this condition implies, for arbitrary $\epsilon^0$ and $N$, the equation
\begin{eqnarray}
    P^{(T)} \frac{{\rm d} H^{\rm matter}}{{\rm d} (\partial_b Z^i)} - P^{(Z)}_i \frac{{\rm d} H^{\rm matter}}{{\rm d} (\partial_b T)}
    \bigg|_{\text{O.S.}} = 0
    \ ,
    \label{eq:Covariance conditon in Poisson brackets - spatial component - ansatz}
\end{eqnarray}
where ${\rm d}$ stand for total derivative.
The solution to this equation implies that the phase-space dependence on the configuration variables' derivatives of the constraint ansatz is of the form $H_s^{\rm matter} \left( P^{(T)} \partial_a T + P^{(Z)}_i \partial_a Z^i \right)$.
Consistency between this solution to the covariance condition, and the solution to the normalization condition (\ref{eq:Constraint ansatz - symmetry, normalized}), yields
\begin{eqnarray}
    H^{\rm matter}
    =
    H_s^{\rm dust}
    - \bar{f} \left( (P^{(Z)})^2 / (P^{(T)})^2 \right)
    \ .
    \label{eq:Constraint ansatz - symmetry generator, normalized, vector covariance}
\end{eqnarray}

We now focus on the time component of the covariance condition $\delta_\epsilon u_t |_{\text{O.S.}} = \mathcal{L}_\xi u_t |_{\text{O.S.}}$ which, using (\ref{eq:Off-shell gauge transformations for lapse}), (\ref{eq:Off-shell gauge transformations for shift}), (\ref{eq:Normalization condition - Time component of velocity}), (\ref{eq:r.h.s covariance conditon - time component}), and (\ref{eq:Covariance conditon in Poisson brackets - spatial component}), becomes
\begin{equation}
    \frac{1}{\epsilon^0} \left\{ \sqrt{s + q^{a b} u_b u_c} , H [\epsilon^0] \right\}
    - u_a q^{a b} \frac{\partial_b \epsilon^0}{\epsilon^0}
    \bigg|_{\text{O.S.}}
    =
    \frac{1}{N} \left\{ \sqrt{s + q^{a b} u_b u_c} , H [N] \right\}
    - u_a q^{a b} \frac{\partial_b N}{N}
    \bigg|_{\text{O.S.}}
    \ .
\end{equation}
Using the spacetime covariance condition (\ref{eq:Covariance condition of 3-metric - reduced}) this further simplifies to
\begin{equation}
    \frac{u_a q^{a b}}{\epsilon^0} \left( \frac{P^{(T)}}{H^{\rm dust}_s} \{ u_a , H [\epsilon^0]\}
    - \partial_b \epsilon^0 \right)
    \bigg|_{\text{O.S.}}
    =
    \frac{u_a q^{a b}}{N} \left( \frac{P^{(T)}}{H^{\rm dust}_s} \{ u_a , H [N]\}
    - \partial_b N \right)
    \bigg|_{\text{O.S.}}
    \ .
    \label{eq:Covariance conditon in Poisson brackets - time component}
\end{equation}
If the spatial covariance condition (\ref{eq:Covariance conditon in Poisson brackets - spatial component}) is satisifed, then (\ref{eq:Covariance conditon in Poisson brackets - time component}) is automatically satisfied too, thus, it does not imply an independent equation.


\subsection{Anomaly-freedom and spacetime covariance}
\label{sec:Anomaly-freedom and spacetime covariance}

Note that the undetermined function $\bar{f}$ in (\ref{eq:Constraint ansatz - symmetry generator, normalized, vector covariance}) can still depend on the structure function, which we have suppressed so far for notational ease.
This dependence can be fully addressed by the requirement of anomaly-freedom.

Anomaly-freedom is realized if the Hamiltonian constraint satisfies the brackets (\ref{eq:Hypersurface deformation algebra - HaHa})-(\ref{eq:Hypersurface deformation algebra - HH}).
The bracket (\ref{eq:Hypersurface deformation algebra - HaHa}) is trivially satisfied because we have assumed the classical form of the diffeomorphism constraint.
For the bracket (\ref{eq:Hypersurface deformation algebra - HHa}) to be satisfied, the Hamiltonian constraint must be a density-weight-one function.
We assume this is the case for the gravitational contribution.
On the other hand, imposing that the matter contribution of the form (\ref{eq:Constraint ansatz - symmetry generator, normalized, vector covariance}) is a density-weight-one function fixes its dependence on the structure function,
\begin{eqnarray}
    \bar{f} &=&
    \sqrt{\det q} P \left( (P^{(Z)})^2/(P^{(T)})^2 \right)
    \ ,
\end{eqnarray}
where $P$ is an undetermined function of $\delta^{ij} W_i W_j$, and may also depend on the gravitational variables as long as it remains a density-weight-zero.
We will discuss the physical meaning of the function $P$ below and conclude that it in fact plays the role of pressure.
For now, we simply note that $P$ is a function of $W^2 = (P^{(Z)})^2 / (P^{(T)})^2$, the magnitude of the internal velocity of the fluid's particles.
This is precisely one of the basic ways to understand pressure, as an effect of particle flow.

We now focus on the spacetime covariance condition (\ref{eq:Covariance condition of 3-metric - reduced}) and the last bracket (\ref{eq:Hypersurface deformation algebra - HH}), which is the most complicated one.
We first note that since $H^{\rm grav}$ is independent of the fluid's variables as found in the previous conditions, it follows that the gravitational contribution must already satisfy $\{ H^{\rm grav} [ N_1 ] , H^{\rm grav} [ N_2 ] \} = - \vec{H}^{\rm grav} [ q^{a b} ( N_2 \partial_b N_1 - N_1 \partial_b N_2 )]$ by its own, and this in turn implies that $q^{a b}$ is independent of the fluid's variables.
Since the covariance condition for the structure function (\ref{eq:Covariance condition of 3-metric - reduced}) involves the full constraint $H=H^{\rm grav}+H_s^{\rm matter}$, but both $q^{ab}$ and $H^{\rm grav}$ are independent of the fluid's variables, it follows that the equation (\ref{eq:Covariance condition of 3-metric - reduced}) must hold if one replaces $H$ by either of the two contributions independently.
We assume that the gravitational contribution already satisfies this covariance condition.
On the other hand, the fluid's contribution to this condition becomes
\begin{equation}
    \frac{\partial \{ q^{a b} , P [\sqrt{\det q} \epsilon^0] \}_{\rm grav}}{\partial (\partial_c \epsilon^0)} \bigg|_{\text{O.S.}}
    = \frac{\partial \{ q^{a b} , P [\sqrt{\det q} \epsilon^0] \}_{\rm grav}}{\partial (\partial_c \partial_d \epsilon^0)} \bigg|_{\text{O.S.}}
    = \dotsi
    = 0
    \ ,
    \label{eq:Covariance condition of 3-metric - modified pressure}
\end{equation}
where we use the subscript 'grav' to restrict the action of the Poisson brackets to the gravitational variables.

Coming back to the bracket (\ref{eq:Hypersurface deformation algebra - HH}), a direct evaluation shows that
\begin{equation}
    \{ H_s^{\rm matter} [ N_1 ] , H_s^{\rm matter} [ N_2 ] \}_{\rm matter}
    = - \vec{H}^{\rm matter} [ q^{a b} ( N_2 \partial_b N_1 - N_1 \partial_b N_2 )]
    \ ,
\end{equation}
where the subscript 'matter' restricts the action of the Poisson brackets to the matter variables.
Furthermore, using the spacetime covariance condition, we find that $\{H^{\rm grav} [N_1] , H_s^{\rm matter} [N_2] \}=-\{H^{\rm grav} [N_1] , P [\sqrt{\det q} N_2]\}_{\rm grav}$.
Therefore, for the bracket (\ref{eq:Hypersurface deformation algebra - HH}) to hold for the full Hamiltonian constraint we obtain the following condition on the modified pressure function,
\begin{equation}
    \{ P [\sqrt{\det q} N_1] , H [N_2] \}
    - \{P [\sqrt{\det q} N_2] , H [N_1]\}
    =
    \{P [\sqrt{\det q} N_2] , P [\sqrt{\det q} N_1]\}
    \ .
    \label{eq:Anomaly-freedom condition on modified pressure}
\end{equation}
The two conditions (\ref{eq:Covariance condition of 3-metric - modified pressure}) and (\ref{eq:Anomaly-freedom condition on modified pressure}) are nontrivial and there may be room for the pressure function to receive modifications\textemdash note that the action of the Poisson brackets on the structure function in (\ref{eq:Anomaly-freedom condition on modified pressure}) always vanishes because the spacetime covariance condition requires that the its bracket with $H^{\rm grav}$ and $P$ does not generate spatial derivatives of the $N_1$ and $N_2$, while the non-derivative terms cancel due to the antisymmetry of the bracket.


\subsection{Pressure regained}

Generally, in EMG the metric no longer enters the Hamiltonian directly as a phase space variable and, therefore, Einstein's equations no longer hold in the sense that a reconstructed Lagrangian is not necessarily minimized with respect to variations of the metric and one does not recover Einstein's equations except in the classical limit.
However, when the matter Hamiltonian resembles the classical one, then the same classical stress-energy components are obtained \textemdash up to the emergent metric.
This is the case for the constraint (\ref{eq:Constraint ansatz - symmetry, full covariance, normalized, anomaly-free}) when the $P$ function is independent of the gravitational variables (which we assume in this subsection to extract its physical meaning), and this will help us in interpreting the variables of the modified theory with the more familiar stress-energy components.
The following canonical stress-energy components are the values that Eulerian observers would measure, that is, in the foliation basis for which we will use the components
\begin{equation}
    u_0 \equiv n^\mu u_\mu
    = - \frac{\sqrt{s (P^{(T)})^2 + q^{a b} H^{\text{matter}}_a H^{\text{matter}}_b}}{P^{(T)}}
    \ ,
\end{equation}
\begin{equation}
    u_a = - \frac{H_{a}^{\text{matter}}}{P^{(T)}}
\end{equation}
\begin{eqnarray}
    g_{0 0} \equiv n^\mu n^\nu g_{\mu \nu} = -1
    \ .
\end{eqnarray}

The energy density is
\begin{eqnarray}
    \rho^{\rm (E)} \equiv \frac{1}{\sqrt{\det q}} \frac{\delta H_s^{\rm matter} [N]}{\delta N}
    =
    \rho_s^{\rm dust} u_0 u_0
    - P
    \ ,
\end{eqnarray}
where the expression for $\rho_s^{\rm dust}$ is given by (\ref{eq:Dust energy density}) by replacing the classical structure function with the emergent one.
The mass current density is
\begin{eqnarray}
    J_a^{\rm (E)} \equiv \frac{1}{\sqrt{\det q}} \frac{\delta \vec{H}^{\rm matter} [\vec{N}]}{\delta N^a}
    = - \rho_s^{\rm dust} u_0 u_a
    \ ,
\end{eqnarray}
and the spatial stress tensor is
\begin{eqnarray}
    S_{a b}^{\rm (E)} \equiv \frac{2}{N \sqrt{\det q}} \frac{\delta H_s^{\rm matter} [N]}{\delta q^{a b}}
    = \rho_s^{\rm dust} u_a u_b
    + q_{a b} P
    \ .
\end{eqnarray}
The pressure is
\begin{eqnarray}
    P^{\rm (E)} \equiv \frac{1}{3} q^{a b} S_{a b}
    = \frac{1}{3} \rho_s^{\rm dust} q^{a b} u_a u_b
    + P
    \ .
\end{eqnarray}

We now compare the canonical energy results to the respective components of the stress-energy tensor $T_{\mu \nu} = \rho_s u_\mu u_\nu + P (g_{\mu \nu} + u_\mu u_\nu)$.
The normal component is
\begin{eqnarray}
    T_{0 0} \equiv n^\mu n^\nu T_{\mu \nu}
    = \left( \rho_s + P \right) u_0 u_0
    - P
    \ .
\end{eqnarray}
A comparison with the canonical energy density, equating $T_{0 0} = \rho^{\rm (E)}_s$, then establishes
\begin{eqnarray}
    \rho_s = \rho_s^{\rm dust} - P
    \ ,
    \label{eq:Energy density in dust frame}
\end{eqnarray}
and confirms $P$ as the pressure function.
This result shows that the energy has a kinetic contribution $\rho_s^{\rm dust}$ from the fluid particles and an interaction contribution $P$ from the pressure.
The normal-spatial components,
\begin{equation}
    T_{0 a} \equiv n^\mu s^\nu_a T_{\mu \nu}
    = - (\rho_s + P) u_0 u_a
    = - \rho_s^{\rm dust} u_0 u_a
    \ ,
\end{equation}
then agree with the Eulerian mass current upon equating $T_{0 a} = J_a^{\rm (E)}$.
Finally, the spatial components
\begin{equation}
    T_{a b} = ( \rho_s + P) u_a u_b + q_{a b} P
    = \rho_s^{\rm dust} u_a u_b
    + q_{a b} P
    \ ,
\end{equation}
also agree with the canonical stress $T_{a b} = S_{a b}^{\rm (E)}$.
The pressure and the freedom of the equation of state have been, thus, regained.

\subsection{Matter covariance conditions II}

As a final consistency check, we want to make sure that the energy density (\ref{eq:Energy density in dust frame}) is a spacetime scalar, which is equivalent to the statement that $\rho_s^{\rm dust}$ and $P$ are spacetime scalars, that is, that their gauge transformations correspond to the infinitesimal coordinate transformation of a scalar field.
This is the case if and only if the equation $\delta_\epsilon \rho_s^{\rm dust} |_{\text{O.S.}} = \mathcal{L}_\xi \rho_s^{\rm dust} |_{\text{O.S.}}$, and similarly for $P$, is satisfied.
Performing the ADM decomposition of this equation, and using Hamilton's equations of motion, it can be simplified to
\begin{eqnarray}
    \frac{1}{\epsilon^0} \{ \rho_s^{\rm dust} , H [\epsilon^0] \} \big|_{\text{O.S.}} =
    \frac{1}{N} \{ \rho_s^{\rm dust} , H [N] \} \big|_{\text{O.S.}}
    \ .
    \label{eq:Covariance condition - energy density}
\end{eqnarray}
Using the spacetime and matter covariance conditions, (\ref{eq:Covariance condition of 3-metric - reduced}) and (\ref{eq:Covariance conditon in Poisson brackets - spatial component}), and the fact that $\rho^{\rm dust}_s$ is independent of the gravitational variables, this can be written as
\begin{equation}
    \frac{1}{\epsilon^0} \left( \{ P^{(T)} , H_s^{\rm matter} [\epsilon^0] \}
    - \frac{P^{(T)}}{H^{\rm dust}_s} q^{a b} H^{\rm matter}_a \partial_b \epsilon^0 \right)
    \bigg|_{\text{O.S.}}
    \notag\\
    =
    \frac{1}{N} \left( \{ P^{(T)} , H_s^{\rm matter} [N] \}
    - \frac{P^{(T)}}{H^{\rm dust}_s} q^{a b} H^{\rm matter}_a \partial_b N \right)
    \bigg|_{\text{O.S.}}
    \ .
\end{equation}
This equation is automatically satisfied by the constraint (\ref{eq:Constraint ansatz - symmetry, full covariance, normalized, anomaly-free}).
Similarly, it the covariance condition for the pressure $P$ to transform as a spacetime scalar is obtained from (\ref{eq:Covariance condition - energy density}) by replacing $\rho^{\rm dust}_s$ with $P$.
This condition then reduces to
\begin{equation}
    \frac{\partial \left(\{ P , H[\sqrt{\det q} \epsilon^0] \}\right)}{\partial (\partial_c \epsilon^0)} \bigg|_{\text{O.S.}}
    = \frac{\partial \{ P , H [\sqrt{\det q} \epsilon^0] \}}{\partial (\partial_c \partial_d \epsilon^0)} \bigg|_{\text{O.S.}}
    = \dotsi
    = 0
    \ ,
    \label{eq:Covariance condition of modified pressure}
\end{equation}
where we have used (\ref{eq:Covariance condition of 3-metric - modified pressure}).
The anomaly-freedom condition (\ref{eq:Anomaly-freedom condition on modified pressure}) and the pressure covariance condition (\ref{eq:Covariance condition of modified pressure}) can be combined and respectively simplified to
\begin{equation}
    \{P [\sqrt{\det q} N_2] , P [\sqrt{\det q} N_1]\}
    = 0
    \ .
    \label{eq:Anomaly-freedom condition on modified pressure - reduced by covariance of pressure}
\end{equation}
\begin{equation}
    \frac{\partial \{ P , H^{\rm grav}[\sqrt{\det q} \epsilon^0] \}}{\partial (\partial_c \epsilon^0)} \bigg|_{\text{O.S.}}
    = \frac{\partial \{ P , H^{\rm grav} [\sqrt{\det q} \epsilon^0] \}}{\partial (\partial_c \partial_d \epsilon^0)} \bigg|_{\text{O.S.}}
    = \dotsi
    = 0
    \ ,
    \label{eq:Covariance condition of modified pressure - simplified}
\end{equation}

Therefore, the conditions for symmetry, normalization, anomaly-freedom, and covariance of the 4-velocity imply the covariance of the energy density, but the pressure requires the additional condition (\ref{eq:Covariance condition of modified pressure}) to transform as a spacetime scalar.
When the pressure function is independent of the gravitational variables, this condition is automatically satisfied.

This completes the implementation of all the symmetry, anomaly-freedom, and covariance conditions on the fluid's Hamiltonian constraint contribution, which acquires the form
\begin{equation}
    H_s^{\rm matter}
    =
    H_s^{\rm dust}
    - \sqrt{\det{q}} P
    \ ,
    \label{eq:Constraint ansatz - symmetry, full covariance, normalized, anomaly-free}
\end{equation}
where $P$ is an undetermined function of $\delta^{ij} W_i W_j$ and the gravitational variables such that it preserves its density-weight-zero and satisfies the conditions (\ref{eq:Covariance condition of 3-metric - modified pressure}), (\ref{eq:Anomaly-freedom condition on modified pressure - reduced by covariance of pressure}), and (\ref{eq:Covariance condition of modified pressure - simplified}).
An evaluation of these conditions requires specifying the gravitational phase-space and constraint contribution, which we cannot do in the general case, but we will be able to evaluate them in the spherically symmetric case.

The timelike case $s=1$ of the Hamiltonian constraint for the perfect fluid (\ref{eq:Constraint ansatz - symmetry, full covariance, normalized, anomaly-free}) is the same constraint\textemdash up to the emergent metric and the pressure's dependence on the gravitational variables\textemdash obtained via the ADM decomposition of the action of the perfect fluid \cite{brown1993action}, but does not require the introduction of entropy as an extra phase-space variable, and unlike in this reference the $P$ function we derived does not depend solely on the number density (\ref{eq:Eulerian particle number density - dust}).

\section{Emergent modified gravity: Spherical symmetry}
\label{sec:Spherically symmetric sector}

\subsection{Vacuum}

In the spherically symmetric theory in vacuum, the spacetime metric takes the general form
\begin{equation}
    {\rm d} s^2 = - N^2 {\rm d} t^2 + q_{x x} ( {\rm d} x + N^x {\rm d} t )^2 + q_{\vartheta \vartheta} {\rm d} \Omega^2
    \label{eq:ADM line element - spherical}
    \ .
\end{equation}
The classical spatial metric components can be written in terms of the classical radial and angular densitized triads $E^x$ and $E^\varphi$, respectively,  $q_{xx} = (E^\varphi)^2/E^x$ and $q_{\vartheta \vartheta} = E^x$.

The symplectic structure of the canonical theory is
\begin{equation}
    \{ K_x (x) , E^x (y)\}
    = \{ K_\varphi (x) , E^\varphi (y) \}
    = \delta (x-y)
    \ ,
\end{equation}
where $K_x$ and $K_\varphi$ are the radial and angular components of the extrinsic curvature.
Within spherical symmetry, only the Hamiltonian constraint and the radial diffeomorphism constraint are non-trivial.
Consequently, the hypersurface deformation algebra (\ref{eq:Hypersurface deformation algebra - HaHa})-(\ref{eq:Hypersurface deformation algebra - HH}) simplifies to
\begin{eqnarray}
    \{ H_x [ N_1^x] , H_x [ N^x_2 ] \} &=& - H_x [\mathcal{L}_{N^x_2} N^x_1]
    \ ,
    \label{eq:Hypersurface deformation algebra - HaHa - spherical}
    \\
    \{ H [ N ] , H_x [ N^x]\} &=& - H [ N^x N' ]
    \label{eq:Hypersurface deformation algebra - HHa - spherical}
    \ , \\
    \{ H [ N_1 ] , H [ N_2 ] \} &=& - H_x [ q^{xx} ( N_2 N_1' - N_1 N_2' )]
    \ , \ \
    \label{eq:Hypersurface deformation algebra - HH - spherical}
\end{eqnarray}
where the prime denotes a derivative with respect to the radial coordinate $x$, and $q^{xx}=E^x/(E^\varphi)^2$ is simply the inverse of the classical spatial metric component $q_{xx}$.
The algebra (\ref{eq:Hypersurface deformation algebra - HaHa - spherical})-(\ref{eq:Hypersurface deformation algebra - HH - spherical}) holds even in the presence of matter\textemdash when gauge fields are present, typically the third bracket receives a contribution from the respective Gauss constraint, but we shall neglect this in the following since the perfect fluid is not a gauge field.
We denote the constraints in vacuum by $H^{\rm grav}$ and $H^{\rm grav}_x$.

In emergent modified gravity we keep $H_x^{\rm grav}$ in its classical form
\begin{equation}
    H_x^{\rm grav} =
    E^\varphi K_\varphi'
    - K_x (E^x)'
    \ ,
    \label{eq:Diffeomorphism constraint - Gravity - spherical}
\end{equation}
such that the bracket (\ref{eq:Hypersurface deformation algebra - HaHa - spherical}) is unmodified.
However, we allow for the Hamiltonian constraint $H^{\rm grav}$ to be non-classical.
According to the regaining procedure of EMG, we demand that the modified $H^{\rm grav}$ is anomaly-free, such that it reproduces the brackets (\ref{eq:Hypersurface deformation algebra - HHa - spherical}) and (\ref{eq:Hypersurface deformation algebra - HH - spherical}) up to a modified $q^{xx}$ such that its phase-space dependence differs from the classical one.
In practice, we start with a general ansatz for $H^{\rm grav}$ as function of the phase-space variables and their derivatives up to second order \cite{alonso2021anomaly,Covariance_regained}.
Imposing the anomaly-freedom of the hypersurface deformation algebra (\ref{eq:Hypersurface deformation algebra - HaHa - spherical})-(\ref{eq:Hypersurface deformation algebra - HH - spherical}) restricts this ansatz severely by providing a set of differential equations it must satisfy, and it determines the expression for $q^{xx}$.
The inverse of this modified structure function is then replaced in the line element (\ref{eq:ADM line element - spherical}).
Unlike $q^{xx}$, the angular component $q^{\vartheta \vartheta}$ cannot be recovered from the hypersurface deformation algebra because the underlying spherical symmetry trivializes the angular diffeomorphism constraints.
Therefore, we simply take the classical value $q^{\vartheta \vartheta}=1/E^x$ and use its inverse in (\ref{eq:ADM line element - spherical}).
This determines the emergent metric.

Once the emergent metric has been identified by imposing anomaly-freedom, we must impose the covariance condition (\ref{eq:Covariance condition of 3-metric - reduced}), which in spherical symmetry reduces to the simpler conditions
\begin{eqnarray}
    \frac{\partial (\delta_{\epsilon^0} q^{xx})}{\partial (\epsilon^0)'} \bigg|_{\text{O.S.}}
    = \frac{\partial (\delta_{\epsilon^0} q^{xx})}{\partial (\epsilon^0)''} \bigg|_{\text{O.S.}}
    = \dotsi
    = 0
    \ , \nonumber \\
    \frac{\partial (\delta_{\epsilon^0} q^{\vartheta \vartheta})}{\partial (\epsilon^0)'} \bigg|_{\text{O.S.}}
    = \frac{\partial (\delta_{\epsilon^0} q^{\vartheta \vartheta})}{\partial (\epsilon^0)''} \bigg|_{\text{O.S.}}
    = \dotsi
    = 0
    \ .
    \label{eq:Covariance condition of 3-metric - reduced - spherical}
\end{eqnarray}

The resulting equations of the anomaly-freedom and covariance conditions are restrictive enough that they can all be solved exactly in the vacuum for the most general constraint ansatz where the phase-space dependence involves up to second-order derivatives; the most general vacuum Hamiltonian constraint in spherical symmetry is given by \cite{Covariance_regained}
\begin{eqnarray}
    H^{\rm grav} &=&
    - \lambda_0 \frac{\sqrt{E^x}}{2} \Bigg[ E^\varphi \Bigg(
    - \Lambda_0
    + \frac{\alpha_0}{E^x}
    + \left( c_f \frac{\alpha_2}{E^x}
    + 2 \frac{\partial c_{f}}{\partial E^x} \right) \frac{\sin^2 \left(\bar{\lambda} K_\varphi\right)}{\bar{\lambda}^2}
    \nonumber\\
    &&
    + 2 \left( q \frac{\alpha_2}{E^x} + 2 \frac{\partial q}{\partial E^x} \right) \frac{\sin \left(2 \bar{\lambda} K_\varphi\right)}{2 \bar{\lambda}}
    \Bigg)
    + 4 K_x \left(c_f \frac{\sin (2 \bar{\lambda} K_\varphi)}{2 \bar{\lambda}}
    + q \cos(2 \bar{\lambda} K_\varphi)\right)
    \nonumber\\
    &&
    + \frac{((E^x)')^2}{E^\varphi} \left(
    - \frac{\alpha_2}{4 E^x} \cos^2 \left( \bar{\lambda} K_\varphi \right)
    + \frac{K_x}{E^\varphi} \bar{\lambda}^2 \frac{\sin \left(2 \bar{\lambda} K_\varphi \right)}{2 \bar{\lambda}}
    \right)
    + \left( \frac{(E^x)' (E^\varphi)'}{(E^\varphi)^2}
    - \frac{(E^x)''}{E^\varphi} \right) \cos^2 \left( \bar{\lambda} K_\varphi \right)
    \Bigg]
    \ ,
    \label{eq:Hamiltonian constraint - Gravity - spherical}
\end{eqnarray}
with structure function
\begin{equation}
    q^{x x}
    =
    \left(
    \left( c_{f}
    + \left(\frac{\bar{\lambda} (E^x)'}{2 E^\varphi} \right)^2 \right) \cos^2 \left(\lambda K_\varphi\right)
    - 2 q \bar{\lambda}^2 \frac{\sin \left(2 \bar{\lambda} K_\varphi\right)}{2 \bar{\lambda}}\right)
    \lambda_0^2 \frac{E^x}{(E^\varphi)^2}
    \ .
    \label{eq:Structure function - modified - spherical}
\end{equation}
Furthermore, it can be shown that the constraint (\ref{eq:Hamiltonian constraint - Gravity - spherical}) has the weak observable
\begin{eqnarray}
    \mathcal{M} &=&
    d_0
    + \frac{d_2}{2} \left(\exp \int {\rm d} E^x \ \frac{\alpha_2}{2 E^x}\right)
    \left(
    c_f \frac{\sin^2\left(\bar{\lambda} K_{\varphi}\right)}{\bar{\lambda}^2}
    + 2 q \frac{\sin \left(2 \bar{\lambda}  K_{\varphi}\right)}{2 \bar{\lambda}}
    - \cos^2 (\bar{\lambda} K_\varphi) \left(\frac{(E^x)'}{2 E^\varphi}\right)^2
    \right)
    \nonumber\\
    &&
    + \frac{d_2}{4} \int {\rm d} E^x \ \left( \left(
    c_{f0}
    + \frac{\alpha_0}{E^x}
    \right) \exp \int {\rm d} E^x \ \frac{\alpha_2}{2 E^x}\right)
    \ ,
    \label{eq:Gravitational weak observable}
\end{eqnarray}
such that $\dot{\mathcal{M}} = \mathcal{D}_H H + \mathcal{D}_x H_x |_{\rm O.S.} = 0$, where $\mathcal{D}_H$ and $\mathcal{D}_x$ are functions of the phase-space, meaning that $\mathcal{M}$ is a constant of the motion.
The parameters $\bar{\lambda}$, $d_0$, and $d_2$ are constants, while the rest of the parameters are functions of $E^x$, all undetermined by the covariance conditions, representing allowed covariant modifications.
The classical constraint and structure function are recovered in the limit $\lambda_0 , c_{f} , \alpha_0 , \alpha_2 \to 1$, and $\bar{\lambda} , q \to 0$, and $\Lambda_0 \to - \Lambda$ if there is a cosmological constant\footnote{An equivalent constraint was recently obtained in \cite{alonsobardaji2023spacetime} starting with a simpler ansatz and hence a special case of the one in \cite{Covariance_regained}.}.
In the classical limit, and further taking $d_0\to0$ and $d_2\to 1$, the observable (\ref{eq:Gravitational weak observable}) becomes the classical mass function.
The parameters $c_f$, $q$, $\lambda_0$, and $\bar{\lambda}$ are characteristic of EMG because they appear directly in the structure function and hence the emergent metric.
On the other hand, the parameters $\Lambda_0$, $\alpha_0$, and $\alpha_2$ are modification functions allowed even for a non-emergent metric, and hence may be studied in the context of 2D dilaton gravity; these modifications have been treated before in the Hamiltonian formalism for instance in \cite{Tibrewala_Midisuperspace,Bojowald_DeformedGR,Tibrewala_Inhomogeneities,Tibrewala_SecondD}, and a Lagrangian approach with similar modifications can be found in \cite{Kunstatter_New2Ddilaton}.

The parameter $\bar{\lambda}$ is special, as it is responsible for the non-singular black hole solution obtained \cite{alonso2022nonsingular} that we previously discussed.
Furthermore, it has an interpretation within LQG as a quantum parameter as we now briefly explain.
The starting point of the loop quantization is carried out by quantizing not the classical phase-space directly, but instead one works in the holonomy-flux representation.
In spherically symmetric LQG \cite{bojowald2000symmetry,bojowald2004spherically}, the holonomies are given by
\begin{eqnarray}
    h^x_e [K_x] &=& \exp \left( i \int_e {\rm d} x\ K_x \right)
    \ , \label{eq:Radial holonomy}
    \\
    h^\varphi_{v, \lambda} [K_\varphi] &=& \exp \left( i \int_\lambda {\rm d} \theta\ K_\varphi \right)
    = \exp \left( i \lambda K_\varphi (v) \right)
    , \ \
    \label{eq:Angular holonomy}
\end{eqnarray}
where $e$ stands for an arbitrary radial curve of finite coordinate length, while $v$ stands for an arbitrary point in the radial line, and $\lambda$ is the coordinate length of an arbitrary angular curve on the 2-sphere intersecting the point $v$.
The explicit integration in the angular holonomy (\ref{eq:Angular holonomy}) is possible due to spherical symmetry, but the radial holonomy integration must remain formal.
Similarly, the fluxes are given by direct integration of the densitized triads $E^x$ and $E^\varphi$ over finite, 2-dimensional surfaces with normals in the radial and angular direction, respectively.
For the following argument, we need only the expression of the holonomies.

Because the loop quantization is based on the holonomy-flux variables, the Hamiltonian constraint must be rewritten in terms of them, rather than the bare curvatures $K_x$ and $K_\varphi$, leading to a modified constraint in the spirit of EMG.
However, the radial holonomy (\ref{eq:Radial holonomy}) is essentially non-local, and cannot be studied within EMG, which has only been formulated locally.
On the other hand, the angular holonomies (\ref{eq:Angular holonomy}) are indeed local when restricting the state space to spherically symmetric states.
Therefore, angular holonomy effects have a chance to appear in the local equations of EMG.
Furthermore, since the angular holonomies can be integrated, they become simple complex exponentials of the angular curvature (\ref{eq:Angular holonomy}).
Because the Hamiltonian constraint must be Hermitian as an operator in the quantum theory, or simply real prior to quantization, the holonomy modifications in the Hamiltonian constraint will appear as trigonometric functions of $K_\varphi$.
This is indeed the case of the EMG constraint (\ref{eq:Hamiltonian constraint - Gravity - spherical}).
Therefore, the parameter $\lambda$ can be given the interpretation of an angular holonomy length within the LQG theory, and hence acquire the status of a quantum parameter.

The holonomy parameter in (\ref{eq:Angular holonomy}) may depend on the scale, here given by a dependence on $E^x$.
While the parameter $\bar{\lambda}$ in the EMG constraint (\ref{eq:Hamiltonian constraint - Gravity - spherical}) is constant, a detailed analysis of canonical transformations done in \cite{Covariance_regained} shows that a non-constant $\lambda$ does arise in EMG, but through a canonical transformation it can be traded in for the constant $\bar{\lambda}$ provided that the other undetermined functions acquire non-classical expressions in the case where $\lambda$ is non-constant.
In the following sections, however, we will only consider a constant holonomy parameter for simplicity, and we drop the bar in $\bar{\lambda}$ and write this \emph{constant} parameter as $\lambda$ for notational ease.
We will sometimes refer to the effects of $\lambda$ as quantum effects due to its interpretation in LQG, but note that the appearance of $\lambda \neq 0$ is allowed by the covariance conditions regardless of LQG.

\subsection{The perfect fluid}

The angular components of the momentum variables of the perfect fluid would imply via (\ref{eq:Perfect fluid symmetry generator - SO(3)}) that it has an angular mass flux or angular momentum, which breaks the spherical symmetry.
Therefore, these angular components must vanish and the perfect fluid is left with only two non-trivial canonical pairs,
\begin{equation}
    \{ T (x) , P_T (y) \}
    = \{ X (x) , P_X (y) \}
    = \delta (x-y)
    \ .
\end{equation}

The covariance condition for the pressure function (\ref{eq:Covariance condition of 3-metric - modified pressure}) involving the angular component of the structure function $q^{\vartheta\vartheta}=1/E^x$ immediately requires that $P$ does not depend on spatial derivatives of $K_x$.
Taking the ansatz $P (E^x , K_\varphi , W^2)$, the anomaly-freedom condition (\ref{eq:Anomaly-freedom condition on modified pressure - reduced by covariance of pressure}) and the spacetime covariance condition (\ref{eq:Covariance condition of 3-metric - modified pressure}) are automatically satisfied, while the pressure covariance condition
(\ref{eq:Covariance condition of modified pressure - simplified}) simplifies to
\begin{eqnarray}
    \frac{\partial P}{\partial K_\varphi} \bigg|_{\text{O.S.}}
    = 0
    \ .
\end{eqnarray}
Therefore, the dependence of the modified pressure on the gravitational variables reduces to only $E^x$.

The Hamiltonian constraint contribution of the perfect fluid is given by\footnote{In \cite{alonsobardaji2023spacetime} it was suggested that the constraint contribution of timelike dust in spherical symmetry could be given by $H^{\rm dust} = \sqrt{P_T^2 + q^{xx} P_T^2 (T')^2}$ using the emergent structure function, but no underlying theory for the dust was presented. While our procedure shows that this version is indeed anomaly-free and covariant, it overlooks the radial phase-space variables of the dust which are necessary for the rise of the (modified) pressure function of the perfect fluid.}

\begin{equation}
    H^{\rm matter}_s =
    \sqrt{ s P_T^2 + q^{x x} \left( P_T T' + P_X X' \right)^2}
    - \sqrt{\det{q}} P \left(E^x , \frac{P_X^2}{P_T^2}\right)
    \ ,
    \label{eq:Hamiltonian constraint - PF - spherical}
\end{equation}
and the diffeomorphism constraint contribution by
\begin{equation}
    H_x^{\rm matter} =
    P_T T'
    + P_X X'
    \ ,
    \label{eq:Diffeomorphism constraint - PF - spherical}
\end{equation}
where we the determinant is understood to be integrated in the angular coordinates, $\sqrt{\det q}=4 \pi \sqrt{q_{x x}} E^x$, and $q^{xx}$ is the emergent structure function (\ref{eq:Structure function - modified - spherical}).

The associated Eulerian energy density is
\begin{equation}
    \rho_s^{\rm (E)} = \frac{1}{\sqrt{\det{q}}} \sqrt{s P_T^2 + q^{x x} \left( P_T T' + P_X X' \right)^2}
    \ ,
    \label{eq:Eulerian dust energy density - spherical}
\end{equation}
and the energy density in the dust frame is
\begin{equation}
    \rho_s^{\rm dust} = \frac{1}{\sqrt{\det{q}}} \frac{P_T^2}{\sqrt{s P_T^2 + q^{x x} \left( P_T T' + P_X X' \right)^2}}
    \ .
    \label{eq:Dust energy density dust frame - spherical}
\end{equation}

The full Hamiltonian constraint in the presence of a perfect fluid is given by $H=H^{\rm grav} + H_s^{\rm matter}$, and the full diffeomorphism constraint by $H_x = H_x^{\rm grav} + H_x^{\rm matter}$.

\subsection{Reflection symmetry surface}

A particular characteristic of the constraints coupled to the perfect fluid that will be relevant for our discussion is the following.
When taking the classical value $q=0$ in (\ref{eq:Hamiltonian constraint - Gravity - spherical}), the full constraints $H=H^{\rm grav} + H_s^{\rm matter}$ and $H_x = H_x^{\rm grav} + H_x^{\rm matter}$ are symmetric under the time reversal operation $K_\varphi \to - K_\varphi$, $K_x \to - K_x$, $T \to - T$, $P_X \to - P_X$.
Furthermore, it possesses an additional reflection-symmetry surface at $K_\varphi = - \pi / (2 \lambda)$ in the sense that, defining $K_\varphi = - \pi/ (2\lambda) - \delta$, the Hamiltonian is symmetric under the operation $\delta \to - \delta$, $K_x \to - K_x$, $T \to - T$, $P_X \to - P_X$, which is almost identical to time-reversal.
The meaning of this symmetry is the following.
Given a solution to the equations of motion in the region $\delta < 0$, the solution in region $\delta > 0$ will be a perfect reflection of the solution in $\delta < 0$, with the only difference being the flow of time.
Thus, the solution in the region $\delta > 0$ will look like the time-reversed solution in the region $\delta < 0$.
We will exploit this symmetry in obtaining the global structure of the collapse solution.

\section{Gravitational collapse of timelike dust}
\label{eq:Gravitational collapse of timelike dust}

\subsection{Frame transformations and gauge fixing}
\label{sec:Frame transformations}

Consider a coordinate transformation to a frame described by a one-form field $v_\mu {\rm d} x^\mu = v_t {\rm d} t + v_x {\rm d} x$, and normalized $g^{\mu \nu} v_\mu v_\nu = - s$, where we take $s = 1 , 0 , -1$ for timelike, null, and spacelike frames, respectively.
Using the inverse metric in the ADM decomposition,
\begin{equation}
    g^{\mu \nu} =
    q^{a b} s^\mu_a s^\nu_a
    - \frac{1}{N^2} \left(t^\mu - N^a s^\mu_a\right) \left(t^\nu - N^b s^\nu_b\right)
    \ ,
    \label{eq:Inverse metric}
\end{equation}
the normalization can be written as
\begin{equation}
    - \frac{(v_t)^2}{N^2}
    + 2 \frac{N^x}{N^2} v_t v_x
    + \left(q^{x x} - \frac{(N^x)^2}{N^2}\right) (v_x)^2
    = - s
    \ ,
    \label{eq:Frame normalization}
\end{equation}
which can be solved for $v_t$ in terms of $v_x$ or vice versa, for example,
\begin{equation}
    v_t
    = \pm \sqrt{N^2 \left( s + q^{x x} (v_x)^2 \right)}
    + N^x v_x
    \ ,
    \label{eq:Frame normalization - in/outgoing}
\end{equation}
where the sign choice corresponds to either an ingoing or an outgoing frame.
It will also be convenient to use the components of the velocity $v^\mu \equiv g^{\mu \nu} v_\nu$:
\begin{eqnarray}
    v^t &=& - \frac{v_t - N^x v_x}{N^2}
    \ , \nonumber\\
    v^x &=& \frac{v_t - N^x v_x}{N^2} N^x
    \ .
\end{eqnarray}

The co-velocity is a one-form and, thus, defines a parameter $\gamma$ along its integral curve via ${\rm d} \gamma \equiv v_\mu {\rm d} x^\mu$.
If the curve is timelike (or null) we can replace the time coordinate in the metric for the parameter $\gamma$ as the new time coordinate as follows.
The infinitesimal time along the integral curve is
\begin{equation}
    {\rm d} t = \frac{1}{v_t} {\rm d} \gamma - \frac{v_x}{v_t} {\rm d} x
    \ .
\end{equation}
Substituting this in a metric of the general form (\ref{eq:ADM line element - spherical}) and rearranging terms we obtain
\begin{eqnarray}
    {\rm d} s^2
    &=&
    \frac{- N^2 + q_{x x} (N^x)^2}{(v_t)^2} {\rm d} \gamma^2
    + \frac{2 N^2 q_{x x}}{(v_t)^2 v_x} \left(
    \frac{N^x}{N^2} v_t v_x
    + \left( q^{x x}
    - \frac{(N^x)^2}{N^2} \right) (v_x)^2
    \right) {\rm d} \gamma {\rm d} x
    \notag\\
    &&- \frac{N^2 q_{x x}}{(v_t)^2}
    \left( - \frac{(v_t)^2}{N^2}
    + 2 \frac{N^x}{N^2} v_t v_x
    + \left( q^{x x}
    - \frac{(N^x)^2}{N^2} \right) (v_x)^2 \right) {\rm d} x^2
    \notag\\
    &=&
    \frac{- N^2 + q_{x x} (N^x)^2}{(v_t)^2} {\rm d} \gamma^2
    - \frac{2 N^2 q_{x x}}{(v_t)^2 v_x} \left(
    s
    - \frac{(v_t)^2}{N^2}
    + \frac{N^x}{N^2} v_t v_x
    \right) {\rm d} \gamma {\rm d} x
    + s \frac{N^2 q_{x x}}{(v_t)^2} {\rm d} x^2
    \ ,
    \label{eq:Metric under time coordinate transformation}
\end{eqnarray}
where we used (\ref{eq:Frame normalization}) to obtain the last line, and we have suppressed the angular part of the metric.
If this frame is timelike, $s=1$, the parameter $\gamma$ is the proper time along the integral curve and we denote it as $\gamma = - \tau$ (the sign is due to the Lorentzian signature); we can then redefine the metric variables compatible with the new coordinates: 
\begin{equation}
    {\rm d} s^2 =:
    - N_{(\tau)}^2 {\rm d} \tau^2
    + q_{x x}^{(\tau)} ( {\rm d} x + N^x_{(\tau)} {\rm d} \tau )^2
    \ ,
\end{equation}
where
\begin{equation}
    q_{x x}^{(\tau)}
    =
    \frac{N^2 q_{x x}}{(v_t)^2}
    \ , \hspace{0.5cm}
    N^x_{(\tau)}
    =
    - v^x
    \ , \hspace{0.5cm}
    N_{(\tau)}
    = 1
    \ .
\label{eq:Observer's frame coordinate transformation}
\end{equation}
Furthermore, in the new coordinates the co-velocity has the components $v_{\mu} {\rm d} x^\mu = - {\rm d} \tau$, and, hence, the velocity components are
\begin{eqnarray}
    v^\tau_{\rm (new)} &=& g^{\tau \mu} v_{\mu} = - g^{\tau \tau} = 1
    \ , \nonumber\\
    v^x_{\rm (new)}
    &=& g^{x \mu} v_\mu = - g^{x \tau}
    = - N_{\tau}^x
    \ .
    \label{eq:Velocities - Observer frame}
\end{eqnarray}
Therefore, the frame of an observer characterized by its normalized time-like co-vector field $v_\mu$ will, in the canonical context, always be given by a unit lapse $N=1$, while a residual gauge freedom rests on the undetermined shift which is the observer's (negative) radial velocity $N^x = - v^x$.
We will exploit this fact in the following sections to obtain the spacetime solutions associated to free-falling observers directly from the canonical equations of motion.

On the other hand, if the frame is null, $s=0$, the parameter $\gamma$ is an affine parameter and we denote it as $\gamma = u$.
In this case, the metric (\ref{eq:Metric under time coordinate transformation}) becomes of the Eddington-Finkelstein form
\begin{equation}
    {\rm d} s^2
    =
    - \frac{N^2 - q_{x x} (N^x)^2}{(v_t)^2} {\rm d} u^2
    + 2 \frac{\sqrt{N^2 q_{x x}}}{v_t} {\rm d} u {\rm d} x
    \ .
    \label{eq:Eddington-Finkelstein metric form}
\end{equation}
This cannot be expressed in the ADM decomposition which is based on spatial foliations, hence, there is no gauge choice in the canonical formalism to reproduce these coordinates, but they are still useful to extend coordinate charts once the spacetime solution is known in other coordinates.

For completion, similar coordinate transformations can be done for spacelike and null frames that replace the spatial coordinate.
In particular, starting from a metric of the general form (\ref{eq:ADM line element - spherical}), two null transformations described by the one-forms ${\rm d} u = v^{(1)}_\mu {\rm d} x^\mu$ and ${\rm d} v = v^{(2)}_\mu {\rm d} x^\mu$ respectively replacing the time and the spatial coordinates to null coordinates $u$ and $v$ renders the metric into a Kruskal-Szekeres form:
\begin{equation}
    {\rm d} s^2
    =
    - \frac{N^2 - q_{x x}(N^x)^2}{v^{(1)}_t v_t^{(2)}} {\rm d} u {\rm d} v
    \ .
    \label{eq:Kruskal-Szekeres metric form}
\end{equation}
The coordinates $(u , v)$ can be more easily related to the original coordinates $(t,x)$ by using (\ref{eq:Frame normalization - in/outgoing}) together with $v_t v^t + v_x v^x = 0$ to obtain
\begin{equation}
    \frac{v_t^{(i)}}{v_x^{(i)}}
    = s_{(i)} \sqrt{N^2 q^{x x}} + N^x
    \ ,
\end{equation}
where $s_{(1)} = + 1$ and $s_{(2)} = - 1$.
Then
\begin{eqnarray}
    {\rm d} \gamma^{(i)}
    &=&
    v_t^{(i)} {\rm d} t + v_x^{(i)} {\rm d} x
    \notag\\
    &=&
    v_t^{(i)} \left( {\rm d} t + \left(s_{(i)} \sqrt{N^2 q^{x x}} + N^x\right)^{-1} {\rm d} x \right)
    \ ,\ \
    \label{eq:Null coordinates - Kruskal-Szekeres}
\end{eqnarray}
where $\gamma^{(1)} = u$ and $\gamma^{(2)} = v$.
If the spacetime in the original coordinates is static, then we can choose null geodesics which have constant $v_t^{(i)}$ (they are Killing conserved quantities) and can be absorbed into $u$ and $v$, and the expression (\ref{eq:Null coordinates - Kruskal-Szekeres}) can be directly integrated.

\subsection{The fluid frame and the Gullstrand-Painlev\'e gauge}

We start by choosing a space foliation compatible with the fluid frame.
This means that the co-velocity of the fluid has the components $u_\mu = n_\mu = g_{\mu \nu} n^\nu$, or $u_t=-N$ and $u_x=0$.
The vanishing of the spatial component then implies that $H_x^{\text{matter}} = 0$.
Furthermore, since the velocity of the fluid has been adapted to the foliation, the fluid variable $T$ can parametrize time, that is, $\dot{T}=1$.
Using the equations of motion for this, combined with $H_x^{\text{matter}} = 0$, $u_t=-N$, and $u_X=0$, implies the equations
\begin{equation}
    N = 1
    \ , \hspace{1cm}
    P_X = 0
    \ .
\end{equation}
The unit lapse is consistent with our previous discussions such that this gauge is indeed associated to the fluid frame.
We will see later that the choice of adapting the space foliation to the fluid frame leads to the Oppenheimer-Snyder model of uniformly distributed dust in the classical case with vanishing pressure \cite{oppenheimer1939continued,kanai2011gravitational}, but, unlike the uniformity conditions, the adaptation of the foliation to the fluid frame can be easily extended to EMG.
Finally, the on-shell conditions $H_x= 0$ and $H=0$ respectively simplify to
\begin{subequations}
\label{eq:Gravitational constraints - dust frame}
\begin{equation}
    K_x = \frac{E^\varphi K_\varphi'}{(E^x)'}
    \ ,
\end{equation}
\begin{equation}
    H^{\text{grav}} = - P_T - \sqrt{\det{q}} P
    \ .
\end{equation}
\end{subequations}

We now define the Gullstrand-Painlev\'e (GP) gauge as
\begin{equation}
    N = 1
    \ , \hspace{1cm}
    E^x = x^2
    \ ,
    \label{eq:Painleve-Gullstrand gague}
\end{equation}
which is compatible with the fluid frame due to the unit lapse.
This choice must satisfy the consistency equation $\dot{E}^x = 0$, which implies the value of the shift,
\begin{equation}
    N^x
    =
    - \lambda _0 \frac{\sin \left(2 \lambda K_{\varphi}\right)}{2 \lambda} \left( c_f + \lambda^2 \frac{x^2}{(E^{\varphi})^2} \right)
    \ .
    \label{eq:Shift - Painleve-Gullstrand}
\end{equation}
Following our previous discussion on frame transformations involving unit lapse, we conclude that the shift (\ref{eq:Shift - Painleve-Gullstrand}) provides the negative observer's radial velocity and, since this is the fluid frame itself, it is the fluid velocity.
The general equation of motion for the remaining fluid variable is
\begin{equation}
    \dot{X}
    =
    N^x X'
    \ .
    \label{eq:Dust X equations of motion - PG gauge}
\end{equation}

The remaining non-trivial equations are given by the equations of motion $\dot{K}_\varphi$ and $\dot{E}^\varphi$.
We will solve these equations for two special cases below, the classical case, and the case with holonomy modifications.

\subsection{Classical collapse}

An exact solution of the classical collapse of dust is known \cite{oppenheimer1939continued,kanai2011gravitational}.
We will reproduce this result with our canonical methods because it will be useful as a guide to later solve the modified equations and to check the classical limit.
To this end, we will take the full classical limit given by $c_f , \lambda_0 , \alpha_i \to 1$, $\lambda, q \to 0$, with vanishing cosmological constant and pressure $\Lambda_0 = P = 0$.

\subsubsection{Star-exterior}

The classical GP equations of motion for dust are
\begin{subequations}
\begin{equation}
    N^x
    =
    - K_{\varphi}
    \ .
    \label{eq:Shift - Painleve-Gullstrand - Classical vacuum}
\end{equation}
\begin{equation}
    K_\varphi \frac{P_T}{E^\varphi}
    + \frac{1}{2} \partial_{t_{\rm GP}} \left( K_\varphi^2 \right)
    + \frac{x^2}{(E^\varphi)^2} \partial_{t_{\rm GP}} \ln E^{\varphi}
    = 0
    \ ,
\end{equation}

\begin{equation}
    \dot{K}_\varphi
    =
    - \frac{1}{2 x}
    \Bigg[
    1
    - \frac{x^2}{(E^\varphi)^2}
    + \left(x K_\varphi^2\right)'
    \Bigg]
    \ ,
\end{equation}
\begin{equation}
    \partial_{t_{\rm GP}} \left(\frac{E^{\varphi}}{x}\right)
    =
    - K_\varphi \left(\frac{E^{\varphi}}{x}\right)'
    \ .
\end{equation}
\label{eq:Equation of motion - Painleve-Gullstrand - Classical}
\end{subequations}

As is well-known, the exterior Schwarzschild metric in vacuum,
\begin{equation}
    {\rm d} s^2 =
    - \left(1-\frac{2 M}{x}\right) {\rm d} t^2
    + \left(1-\frac{2 M}{x}\right)^{-1} {\rm d} x^2
    + x^2 {\rm d} \Omega^2
    \ ,
    \label{eq:Spacetime metric - Schwarzschild gauge - Static}
\end{equation}
is the solution to the equations of motion in the Schwarzschild gauge given by
\begin{equation}
    N^x = 0
    \ , \hspace{1cm}
    E^x = x^2
    \ .
\end{equation}
The (generalized) Gullstrand-Painlev\'e metric,
\begin{equation}
    {\rm d} s^2 = - {\rm d} t_{\rm GP}^2
    + \frac{1}{\varepsilon^2} \left( {\rm d} x - s \sqrt{\frac{2 M}{x} - \frac{2 M}{R_0}} {\rm d} t_{\rm GP} \right)^2
    + x^2 {\rm d} \Omega^2
    ,
    \label{eq:Generalized Painleve-Gullstrand metric - Classical vacuum}
\end{equation}
with constant $\varepsilon$, $s=\pm 1$, and $x < R_0 = 2 M / (1 - \varepsilon^2)$ is obtained either by direct coordinate transformation of the Schwarzschild metric (\ref{eq:Spacetime metric - Schwarzschild gauge - Static}) associated to a frame as explained in Section~\ref{sec:Frame transformations} whose 4-velocity is that of timelike geodesics with Killing-conserved energy $\varepsilon$\textemdash that is, $v_t = - \varepsilon$, i.e., the geodesics are at rest at $x = R_0$\textemdash or by solving the equations of motion in the GP gauge (\ref{eq:Equation of motion - Painleve-Gullstrand - Classical}) for vacuum, that is, for $P_T = 0$.
The signs $s = - 1$ and $s=+1$ are associated to ingoing or outgoing frames, respectively.

For future use, the classical solution to each phase-space variable in the GP gauge is
\begin{eqnarray}
    E^\varphi &=& \frac{x}{\varepsilon}
    \ , \\
    K_\varphi &=& s \sqrt{\frac{2 M}{x} - \frac{2 M}{R_0}}
    \ , \\
    K_x &=& - \frac{s}{2 \varepsilon} \left(\frac{2 M}{x} - \frac{2 M}{R_0}\right)^{-1/2} \frac{M}{x^2}
    \ .
\end{eqnarray}
Substituting these expressions into the vacuum observable (\ref{eq:Gravitational weak observable}), with $d_0=1$ and $d_2$, yields $\mathcal{M}= M$, as expected.

\subsubsection{Star-interior}

We will now obtain the solution for the interior of a star of dust.
Let us assume that the dust at some initial time $t_0$ is at rest confined to the region $x<R_0$, i.e., the star has an initial radius $R_0$.
The radius of the star at other times, $R (t)$, follows a geodesic motion and because the interior metric must match the exterior one at $x = R(t)$, the radius $R(t)$ follows the geodesic equation of the exterior metric; this equation is easily obtained by recalling that in GP coordinates the shift is the negative velocity of the frame $N^x = - {\rm d} x / {\rm d} t_{\rm G P}$:
\begin{eqnarray}
    \frac{{\rm d} R}{{\rm d} t_{\rm G P}} = s \sqrt{\frac{2 M}{R} - \frac{2 M}{R_0}}
    \ ,
\end{eqnarray}
whose general solution is the inversion of
\begin{eqnarray}
    t_{\rm GP} - t_0
    &=&
    \frac{s R}{\varepsilon^2 - 1} \sqrt{\frac{2 M}{R} - \frac{2 M}{R_0}}
    - \frac{s M}{(\varepsilon^2 - 1)^{3/2}} \ln \left( \frac{\sqrt{\frac{2 M}{R_0}} + \sqrt{\frac{2 M}{R} - \frac{2 M}{R_0}}}{\sqrt{\frac{2 M}{R_0}} - \sqrt{\frac{2 M}{R} - \frac{2 M}{R_0}}} \right)
    \ ,
\end{eqnarray}
and the constant $\varepsilon$ is the Killing conserved energy of the free-falling frame at rest at the radius $R_0$.
The case $\varepsilon\to 1$, $s=-1$ corresponds to collapsing dust from rest at infinity such that
\begin{eqnarray}
    R(t_{\rm G P}) = \left(\frac{9 M}{2} t_{\rm G P}^2 \right)^{1/3}
    \ .
    \label{eq:Star radius from infinity - classical collapse}
\end{eqnarray}
Substituting $R(t_{G P})$ into the metric (\ref{eq:Generalized Painleve-Gullstrand metric - Classical vacuum}) then gives the boundary metric that joins the star's exterior and interior metrics.
The latter is obtained by solving the GP gauge equations of motion (\ref{eq:Equation of motion - Painleve-Gullstrand - Classical}).
To this end, based on the vacuum solution, we take the ans\"atze
\begin{eqnarray}
    E^\varphi &=& \frac{x}{\varepsilon (t_{\rm G P},x)}
    \ , \nonumber\\
    K_\varphi &=& s \sqrt{\frac{2 m (t_{\rm G P},x) }{x} + \varepsilon (t_{\rm G P},x)^2 - 1}
    \ ,
    \label{eq:Interior ansatz}
\end{eqnarray}
with $s = \pm 1$.
The intuition behind this ansatz is that at any given time $t_{\rm GP}$ there will be a total mass $m (t_{\rm G P},x)$ contained within the radius $x$, which will be experienced by the observers parametrized by $\varepsilon$, which is no longer a Killing conserved quantity.
Defining $\varepsilon^2 = 1 - E$ and using the ansatz (\ref{eq:Interior ansatz}) the last two equations of (\ref{eq:Equation of motion - Painleve-Gullstrand - Classical}) become
\begin{eqnarray}
    \dot{m}
    &=&
    - K_\varphi m'
    \ , \\
    \dot{E}
    &=&
    - K_\varphi E'
    \ .
\end{eqnarray}
Using the method of separation of variables, i.e., assuming the forms $E = \lambda_E E_t (t_{\rm G P}) E_x (x)$, and $m = \lambda_m m_t (t_{\rm G P}) m_x (x)$ with $\lambda_E$ and $\lambda_m$ constants, and imposing the boundary conditions $m( t_{\rm G P} , x = R(t_{\rm G P})) = M$ and $E(t_0) = 1 - \varepsilon^2 = 2 M / R_0$, one can solve these equations exactly obtaining
\begin{eqnarray}
    m
    &=&
    M \frac{x^3}{R^3}
    \label{eq:Mass function - Classical collapse}
    \ ,
    \\
    E &=& \frac{2 M}{R_0} \frac{x^2}{R^2}
    = \frac{2 m}{x} \frac{R}{R_0}
    \ ,
    \\
    P_T &=& \frac{m'}{\varepsilon}
    \ ,
\end{eqnarray}
with only (\ref{eq:Dust X equations of motion - PG gauge}), the equation for $X$, being difficult to solve exactly except in the limit $\varepsilon \to 1$, $R_0\to \infty$.
In this limit, and imposing the boundary condition $X (t_{\rm GP} , x=R(t_{\rm GP})) = R(t_{\rm GP})$, the solution is given by
\begin{equation}
    X = \left( \frac{\left(9 M t_{\rm GP}^2/2\right)^{1/2} + x^{3/2}}{2} \right)^{2/3}
    \ .
\end{equation}

This solution describes a uniformly distributed dust with energy density $\rho^{\rm dust} (R)$ and therefore corresponds to the Oppenheimer-Snyder collapse model \cite{oppenheimer1939continued}: The associated energy density is
\begin{equation}
    \rho^{\rm dust}
    =
    \frac{\varepsilon}{4 \pi x^2} P_T
    = \frac{m}{4 \pi x^3 / 3}
    \ ,
\end{equation}
and the global conserved charge (\ref{eq:Perfect fluid symmetry generator - total mass}) with $\alpha=\varepsilon$ gives the total mass
\begin{equation}
    Q[\varepsilon] = \int {\rm d} x\ \varepsilon P_T
    = M \ .
\end{equation}
The resulting metric for the star-interior is given by \cite{kanai2011gravitational}
\begin{equation}
    {\rm d} s^2 = - {\rm d} t_{\rm GP}^2
    + \left(1 - \frac{2 m}{x} \frac{R}{R_0}\right)^{-1} \left( {\rm d} x - s \sqrt{\frac{2 m}{x}} \sqrt{1 - \frac{R}{R_0}} {\rm d} t_{\rm GP} \right)^2
    + x^2 {\rm d} \Omega^2
    \ .
    \label{eq:Generalized Painleve-Gullstrand metric - Star interior}
\end{equation}
The physical singularity at $t_{\text{GP}} \to 0$ persists, where $R \to 0$: all of the dust collapses into the coordinate point $(t_{\text{GP}} \to 0 , x \to 0)$, and the usual spacelike singularity of the vacuum black hole appears from then on.

If the star falls from infinity, then $R_0 \to \infty$ and $R(t_{G P})$ is given by (\ref{eq:Star radius from infinity - classical collapse}); the metric (\ref{eq:Generalized Painleve-Gullstrand metric - Star interior}) in this case can be rewritten as
\begin{equation}
    {\rm d} s^2 = - {\rm d} t_{\rm GP}^2
    + \left( {\rm d} x - K_\varphi {\rm d} t_{\rm GP} \right)^2
    + x^2 {\rm d} \Omega^2
    \ ,
    \label{eq:Generalized Painleve-Gullstrand metric in curvature terms - Star interior}
\end{equation}
where the curvature is simply $K_\varphi = - \sqrt{2 m / x}$ with range $K_\varphi \in [0 , - \infty)$ \textemdash corresponding to $t_{\text{GP}} < 0$.
However, the range $K_\varphi \in ( \infty , 0 ]$\textemdash corresponding to $t_{\text{GP}} > 0$\textemdash is a solution too, describing the time-reversed process: an exploding star with the dust coming out of the singularity.

\begin{figure}[h]
    \begin{subfigure}{.5\columnwidth}
        \centering
        \includegraphics[trim=0cm 0cm 0cm 0cm,clip=true,width=\columnwidth]{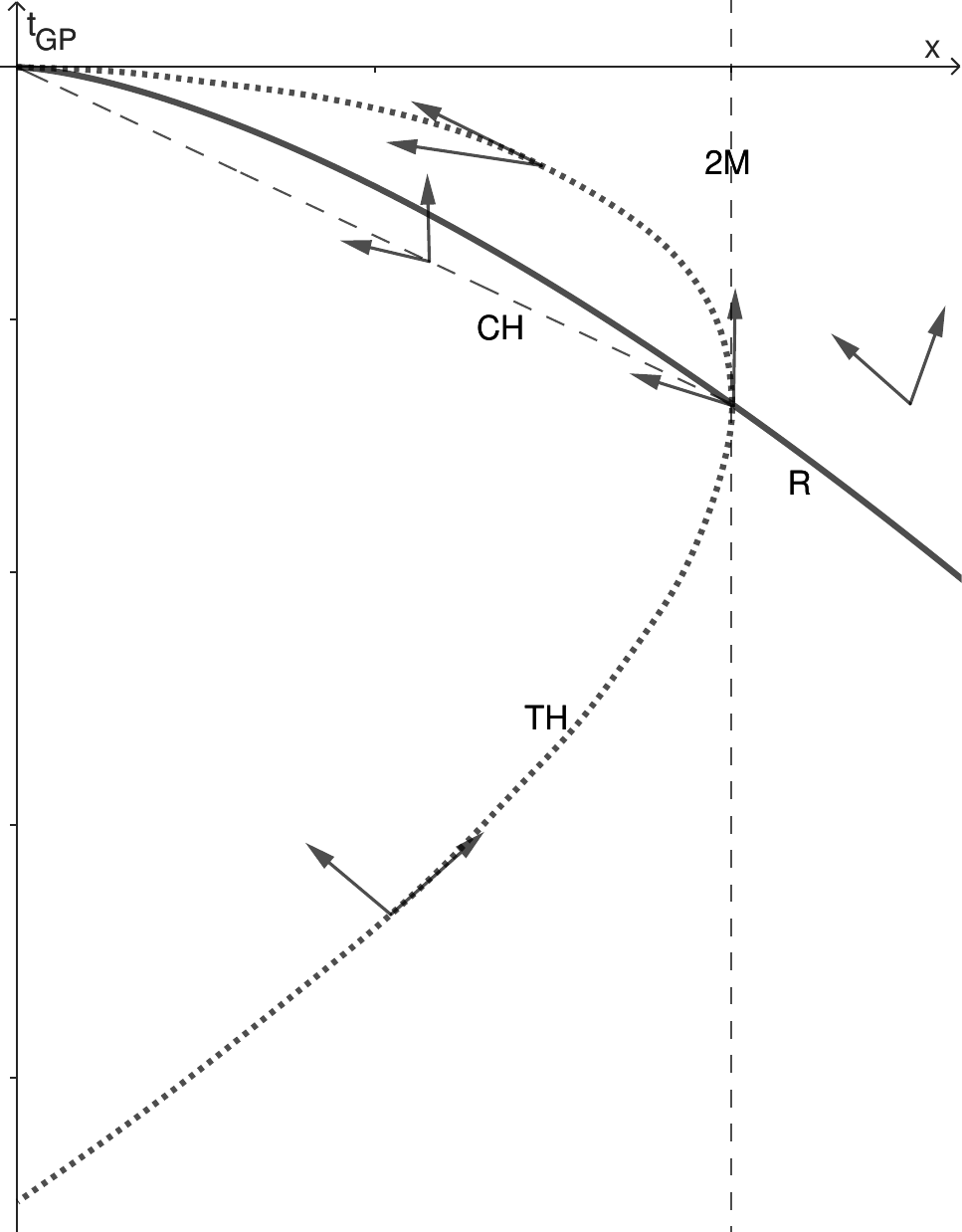}
        \caption{\empty}
        \label{fig:Classical_collapse_GP_a}
    \end{subfigure}%
    \begin{subfigure}{.5\columnwidth}
        \centering
        \includegraphics[trim=14cm 5.1cm 13.5cm 5.9cm,clip=true,width=\columnwidth]{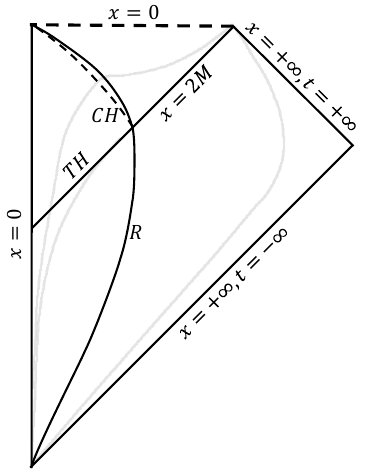}
        \caption{\empty}
        \label{fig:Classical_collapse_GP_b}
    \end{subfigure}
\caption{Classical collapse of a dust star with infinite initial radius $R_0 \to \infty$ in (a) GP coordinates and (b) its conformal diagram.
Some light cones have been represented with arrows in (a) for reference on the causal structure.
    The solid line labeled by $R$ is the star's radius.
    It crosses its own horizon at the time $t_{\rm GP} = - 4 M / 3$.
    The causal horizon, labeled as CH, is defined by the vanishing velocity of outgoing null-rays in GP coordinates and beyond which it changes sign, becoming ingoing null rays, and constant $x$ curves become spacelike.
    This surface is given by the solution to $2 m = x$, or $t_{\rm GP}=- 2 x / 3$.
    The trapping horizon, labeled as $TH$, is the surface beyond which light cannot escape the event horizon $x=2M$ in vacuum.
    The trapping horizon is obtained by solving the outgoing null geodesic that intersects the star's radius at the event horizon.
    Unlike in the vacuum solution, in the presence of dust the trapping and causal horizons are not identical and neither of them is a constant $x$ curve.
    The faint gray lines in (b) denote constant $x$ curves.}
    \label{fig:Classical_collapse_GP}
\end{figure}

The spacetime diagram of the collapse in GP coordinates is shown in Fig.~\ref{fig:Classical_collapse_GP}.

\subsection{Holonomy model collapse}

Having solved the classical system, we are now in the position to solve the modified equations with finite holonomy parameter $\lambda$.
For simplicity, we take the limit $c_f , \alpha_0 , \alpha_2 \to 1$ and consider constant $\lambda_0$ as an aid to rescale the resulting metric.
We also take vanishing cosmological constant and pressure $\Lambda_0 = P = 0$.

\subsubsection{Star-exterior}

Similar to the classical system, the vacuum equations of motion in the Schwarzschild gauge
\begin{equation}
    N^x = 0
    \ , \hspace{1cm}
    E^x = x^2
    \ ,
\end{equation}
can be solved exactly, yielding the emergent metric \cite{alonso2022nonsingular}
\begin{eqnarray}
    {\rm d} s^2 =
    - \left(1-\frac{2 M}{x}\right) \frac{{\rm d} t^2}{\mu^2 \lambda_0^2}
    + \left( 1
    + \lambda^2 \left(1-\frac{2 M}{x}\right) \right)^{-1}
    \left(1-\frac{2 M}{x}\right)^{-1} \frac{{\rm d} x^2}{\lambda_0^2}
    + x^2 {\rm d} \Omega^2
    \ ,
    \label{eq:Spacetime metric - Schwarzschild gauge - Static - Holonomy}
\end{eqnarray}
where $\mu$ is a constant of the equations of motion that, together with $\lambda_0$, may rescale the metric, and $M$ is a constant that can be related to the classical mass as we will show.
Demanding asymptotic flatness fixes $\mu = 1/\lambda_0$ and $\lambda_0 = \sqrt{1+\lambda^2}$.
Unlike the classical Schwarzschild metric, this has a second coordinate singularity at
\begin{equation}
    x_\lambda = \frac{2 M \lambda^2}{1+\lambda^2}
    \ , \label{eq:Minimum radius of vacuum}
\end{equation}
which is not a geometric singularity, but a minimum radius placed at the maximum curvature surface.

The spacetime metric in the GP coordinates can again be obtained by either direct coordinate transformation of (\ref{eq:Spacetime metric - Schwarzschild gauge - Static - Holonomy}) using a free-falling timelike frame or by solving the vacuum equations of motion in the GP gauge (\ref{eq:Painleve-Gullstrand gague}); the resulting phase space variables in this gauge are given by
\begin{eqnarray}
    N^x &=& - s \sqrt{\frac{2 M}{x}
    - \frac{2 M}{R_0}} \sqrt{1
    - \frac{x_\lambda}{x}}
    \ , \\
    E^\varphi &=& \frac{x}{\varepsilon}
    \ , \\
    \frac{\sin^2 (\lambda K_\varphi)}{\lambda^2} &=& \frac{2 M/x + \varepsilon^2 - 1}{1 + \lambda^2 \varepsilon^2}
    \ , \\
    K_x &=& \frac{K_\varphi'}{2 \varepsilon}
    \ .
\end{eqnarray}
Plugging these expressions into the weak observable (\ref{eq:Gravitational weak observable}), with $d_0=0$ and $d_2=1$, yields $\mathcal{M} = M$, reinforcing the interpretation of $M$ as that of the mass even in the modified theory.
The emergent GP metric is
\begin{equation}
    {\rm d} s^2
    =
    - {\rm d} t_{\rm GP}^2
    + \frac{1}{\varepsilon^2} \left( 1-\frac{x_\lambda}{x} \right)^{-1}
    \left({\rm d} x - s \sqrt{\frac{2 M}{x}
    - \frac{2 M}{R_0}} \sqrt{1
    - \frac{x_\lambda}{x}} {\rm d} t_{\rm GP}\right)^2
    + x^2 {\rm d} \Omega^2
    \ ,
    \label{eq:Spacetime metric - GP gauge - Static - Holonomy}
\end{equation}
where the constant $\varepsilon$ is the Killing conserved energy of the free-falling frame, $s = {\rm sgn} (\sin (2 \lambda K_\varphi))$, and
$R_0 = 2 M / (1 - \varepsilon^2)$.

The Schwarzschild time $t$ and the GP time $t_{\rm GP}$ are related by the coordinate transformation
\begin{equation}
    {\rm d} t_{\rm GP}
    =
    \varepsilon {\rm d} t
    - s \sqrt{\frac{2 M}{x} - \frac{2 M}{R_0}}
    \left(1-\frac{2 M}{x}\right)^{-1}
    \sqrt{1 - \frac{x_\lambda}{x}} {\rm d} x
    \ ,
\end{equation}
or
\begin{eqnarray}
    t_{\rm GP}
    &=&
    \varepsilon t
    + s \sqrt{\frac{2 M}{R_0}}
    \Bigg(
    - \sqrt{R_0 - x} \sqrt{x - x_\lambda}
    + 2 \sqrt{R_0-2 M} \sqrt{2 M - x_\lambda} \text{arctanh} \left(\frac{\sqrt{R_0 - 2 M} \sqrt{x - x_\lambda}}{\sqrt{R_0 - x} \sqrt{2M - x_\lambda}}\right)
    \nonumber\\
    &&\qquad
    - \left( R_0 - 4M + x_\lambda\right) \text{arctan} \left(\frac{\sqrt{x-x_\lambda}}{\sqrt{R_0-x}}\right)
    \Bigg)
    \ ,
\end{eqnarray}
where the integration constant has been chosen such that $t_{\rm GP} = 0$ coincides with $t=0$ at $x=x_\lambda$

Using that the shift in the GP gauge is the negative radial velocity of the observers, $N^x = - {\rm d} x / {\rm d} t_{\rm G P}$, the radial coordinate $x=R$ of the timelike geodesics then follows the equation
\begin{equation}
    \frac{{\rm d} R}{{\rm d} t_{\rm G P}}
    =
    s \sqrt{\frac{2 M}{R}
    - \frac{2 M}{R_0}} \sqrt{1 - \frac{x_\lambda}{R}}
    \ ,
    \label{eq:Geodesic velocity - Holonomy collapse - GP}
\end{equation}
and we see that $s=-1$ corresponds to an infalling frame; the acceleration is
\begin{equation}
    \frac{{\rm d}^2 R}{{\rm d} t_{\rm G P}^2}
    =
    \frac{M}{R^2} \left( - \left(\frac{2 M}{R} - \frac{2 M}{R_0}\right)^{-1}
    + \frac{\lambda^2}{1+\lambda^2} \left(1 - \frac{x_\lambda}{R} \right)^{-1} \right) \left(\frac{{\rm d} R}{{\rm d} t_{\rm G P}}\right)^2
    \ .
    \label{eq:Geodesic acceleration - Holonomy collapse - GP}
\end{equation}
The velocity equation (\ref{eq:Geodesic velocity - Holonomy collapse - GP}) has the general solution
\begin{equation}
    t_{\rm GP} - t_0
    =
    \frac{s}{\sqrt{2 M}}
    \Bigg(
    R \sqrt{\frac{R_0}{R}} \sqrt{\frac{R_0}{R} - 1}
    \sqrt{1 - \frac{x_\lambda}{R}}
    + R_0 \left(1 - \frac{x_\lambda}{R_0}\right) \left( \sqrt{\frac{R_0}{R}} \arctan \left(\sqrt{\frac{1 - x_\lambda/x}{R_0/R - 1}}\right)
    - \frac{\pi}{2} \right)
    \Bigg)
    \ ,
\end{equation}
so that the observer is at rest at the radial coordinate $x=R_0$ at time $t_0$.
The geodesic then reaches $R=x_\lambda$ at time
\begin{equation}
    t_{\rm G P}
    =
    t_0
    + \frac{s}{\sqrt{2 M}} \frac{\pi}{2}
    \left( R_0 + x_\lambda\right)
    \ ,
\end{equation}
and at this minimum radius the velocity (\ref{eq:Geodesic velocity - Holonomy collapse - GP}) vanishes, while the acceleration (\ref{eq:Geodesic acceleration - Holonomy collapse - GP}) is finite and positive:
\begin{equation}
    \frac{{\rm d}^2 R}{{\rm d} t_{\rm G P}^2}
    \xrightarrow[R \to x_{\lambda}]{\empty}
    \frac{1}{2 x_{\lambda}} \left( \frac{2 M}{x_{\lambda}}
    - \frac{2 M}{R_0} \right)
    \ .
    \label{eq:Geodesic acceleration at minimum radius - Holonomy collapse - GP}
\end{equation}

If the observer falls from infinity, $R_0 \to \infty$ at $t_{\rm GP} \to - \infty$, the time coordinates are instead related by
\begin{equation}
    t_{\rm G P} =
    t
    - s \sqrt{2 M} \left( 2 \sqrt{x-x_\lambda}
    - \sqrt{2M-x_\lambda} \ln \left( \frac{\sqrt{2M-x_\lambda}+\sqrt{x-x_\lambda}}{\sqrt{2M-x_\lambda}-\sqrt{x-x_\lambda}} \right) \right)
    \ ,
    \label{eq:Coordinate transformation Schwarzschild-GP - holonomy vacuum}
\end{equation}
where again the integration constant has been chosen such that $t_{\rm GP} = 0$ coincides with $t=0$ at $x=x_\lambda$.
The expression for the geodesic, which we relabel as $R (t_{\rm GP})$, is instead the solution to
\begin{equation}
    t_{\rm G P}
    =
    \frac{2 s}{3}
    R \sqrt{\frac{R}{2 M}}
    \sqrt{1 - \frac{x_\lambda}{R}} \left( 1 + 2 \frac{x_\lambda}{R} \right)
    \ ,
    \label{eq:Geodesic at rest at infinity - holonomy effects - GP}
\end{equation}
where the integration constant has been chosen such that $t_{\rm GP}=0$ at $R=x_\lambda$.
The explicit inversion of (\ref{eq:Geodesic at rest at infinity - holonomy effects - GP}) is given by
\begin{eqnarray}
    R (t_{\rm GP})
    &=&
    - x_\lambda
    + \left( \frac{9 M}{4} t_{\rm GP}^2 + x_\lambda^3
    + \sqrt{\left(\frac{9 M}{4} t_{\rm GP}^2\right)^2 + \frac{9 M}{2} t_{\rm GP}^2 x_\lambda^3} \right)^{1/3}
    \nonumber\\
    &&
    + x_\lambda^2 \left( \frac{9 M}{4} t_{\rm GP}^2 + x_\lambda^3
    + \sqrt{\left(\frac{9 M}{4} t_{\rm GP}^2\right)^2 + \frac{9 M}{2} t_{\rm GP}^2 x_\lambda^3} \right)^{-1/3}
    \ .
    \label{eq:Geodesic at rest at infinity - solution - holonomy effects - GP}
\end{eqnarray}
This geodesic coordinate reaches the minimum radius $x_\lambda = 2 M \lambda^2 / (1+\lambda^2)$ at $t_{\rm G P} = 0$ (and $t=0$) with vanishing velocity $\dot{R} (x = x_\lambda) = 0$, and finite, positive acceleration
\begin{equation}
    \frac{{\rm d}^2 R}{{\rm d} t_{\rm G P}^2}
    \xrightarrow[R \to x_{\lambda}]{\empty}
    \frac{M}{x_{\lambda}^2}
    \ .
    \label{eq:Geodesic acceleration at minimum radius - Holonomy collapse from infinity - GP}
\end{equation}

\subsubsection{Star-interior}

We now focus on the star-interior region in the GP gauge (\ref{eq:Painleve-Gullstrand gague}).
Based on the classical solution for the star-interior and on the vacuum solution in the holonomy model, we take the ans\"atze
\begin{eqnarray}
    E^\varphi &=& \frac{x}{\varepsilon (t_{\rm GP},x)}
    \ , \nonumber\\
    \frac{\sin (\lambda K_\varphi)}{\lambda}
    &=&
    \frac{s}{\sqrt{1 + \lambda^2 \varepsilon^2}} \sqrt{\frac{2 m (t_{\rm G P},x)}{x} + E}
    \ ,
    \label{eq:Interior ansatz - dust - holonomy effects - GP}
\end{eqnarray}
where $s = \pm 1$ is the sign of $\sin (\lambda K_\varphi)$, and we defined $\varepsilon^2 =: 1 - E$.
Using this ansatz the relevant equations of motion in the GP gauge become
\begin{equation}
    N^x
    =
    - \frac{\sin \left(2 \lambda K_{\varphi}\right)}{2 \lambda} \frac{1 + \lambda^2 \varepsilon^2}{1+\lambda^2}
    \ ,
\end{equation}
\begin{equation}
    \frac{\varepsilon}{x} P_T
    + \dot{K}_\varphi
    + \frac{\lambda}{2} \cot (\lambda K_\varphi) \frac{\dot{E}}{1 + \lambda^2 \varepsilon^2}
    = 0
    \ ,
    \label{eq:On-shell condition - dust collapse - holonomy modifications - GP gauge}
\end{equation}
\begin{equation}
    \frac{\dot{m}}{1 + \lambda^2 \varepsilon^2}
    =
    - \frac{\sin (2 \lambda K_\varphi)}{2 \lambda} \frac{m'}{\sqrt{1+\lambda^2}}
    \ ,
    \label{eq:Mass equation - dust collapse - holonomy modifications - GP gauge}
\end{equation}
\begin{equation}
    \frac{\dot{E}}{1 + \lambda^2 \varepsilon^2}
    =
    - \frac{\sin (2 \lambda K_\varphi)}{2 \lambda} \frac{E'}{\sqrt{1+\lambda^2}}
    \label{eq:Observer energy equation - dust collapse - holonomy modifications - GP gauge}
    \ .
\end{equation}

Unfortunately, these equations are too complicated to obtain an exact general solution.
However, we can solve exactly for the special case of the dust collapsing from rest at infinity.
Thus, restricting ourselves to the case $E\to 0$, $R_0 \to \infty$, and defining $m (t_{\rm G P},x) = \frac{x}{2} m_1 (x/t_{\rm G P})$, equation (\ref{eq:Observer energy equation - dust collapse - holonomy modifications - GP gauge}) is trivially satisfied, while equation (\ref{eq:Mass equation - dust collapse - holonomy modifications - GP gauge}) becomes
\begin{equation}
    z^2 \frac{\partial m_1}{\partial z}
    =
    s_2 \sqrt{m_1}
    \sqrt{1 - \frac{\lambda^2 m_1}{1 + \lambda^2}} \left( m_1 + z \frac{\partial m_1}{\partial z}\right)
    \ ,
    \label{eq:Mass equation - holonomy collapse}
\end{equation}
where $z = x / t_{\rm G P}$ and $s_2$ is the sign of $\sin (2 \lambda K_\varphi)$, thus, we shall take $s_2 = -1$ for the collapsing region and $s_2 = + 1$ for the time-reversed version.
This equation has a general solution with one constant of integration, which, however, is an extremely long expression.
Imposing the boundary condition $m (x=R(t)) = M$ determines the integration constant of the solution, which simplifies considerably and now reads
\begin{eqnarray}
    m
    &=&
    \frac{|t_{\rm G P}|}{8 \sqrt{3}} \frac{1+\lambda^2}{\lambda^3}
    \Bigg[ - 3 - \lambda^2 \left(3 - 4 \left(\frac{x}{t_{\rm G P}}\right)^2\right)
    \notag\\
    &&\hspace{-1cm}
    + 3 \left( \frac{8 \lambda^3}{\sqrt{27}} \left(\frac{x}{|t_{\rm G P}|}\right)^3 + \sqrt{1+\lambda^2} \sqrt{1 + \lambda^2 \left(2 - 4 \left(\frac{x}{t_{\rm G P}}\right)^2\right) + \lambda^4 \left(1 - 4 \left(\frac{x}{t_{\rm G P}}\right)^2 + \frac{16}{3} \left(\frac{x}{t_{\rm G P}}\right)^4\right)} \right)^{2/3}\Bigg]
    \notag\\
    &&\hspace{-1cm} \times
    \left( \frac{8 \lambda^3}{\sqrt{27}} \left(\frac{x}{|t_{\rm G P}|}\right)^3 + \sqrt{1+\lambda^2} \sqrt{1 + \lambda^2 \left(2 - 4 \left(\frac{x}{t_{\rm G P}}\right)^2\right) + \lambda^4 \left(1 - 4 \left(\frac{x}{t_{\rm G P}}\right)^2 + \frac{16}{3} \left(\frac{x}{t_{\rm G P}}\right)^4\right)} \right)^{-1/3}
    ,\ \ \
    \label{eq:Mass function - GP}
\end{eqnarray}
and also satisfies $m (x\to 0) \to 0$, more on this below.
In the limit $\lambda \to 0$, the mass function recovers its classical form (\ref{eq:Mass function - Classical collapse}).

From equation (\ref{eq:On-shell condition - dust collapse - holonomy modifications - GP gauge}) we obtain the dust momentum
\begin{eqnarray}
    P_T
    &=&
    \frac{m'}{\sqrt{1+\lambda^2}}
    \ .
\end{eqnarray}
The global conserved charge (\ref{eq:Perfect fluid symmetry generator - total mass}) with $\alpha=\sqrt{1+\lambda^2}$ is precisely the total mass of the star,
\begin{eqnarray}
    Q_0 \left[\alpha\right] = \int {\rm d} x\ \sqrt{1+\lambda^2} P_T = m (x) \big|_0^{R} = M
    \ ,
\end{eqnarray}
and the associated dust energy density is
\begin{eqnarray}
    \rho^{\rm dust}
    %
    %
    &=&
    \frac{|\cos (\lambda K_\varphi)|}{\sqrt{1+\lambda^2}}
    \frac{m'}{4 \pi x^2}
    \ ,
\end{eqnarray}
which shows that the dust is not uniformly distributed as in the classical solution.

Only the equation of motion of the dust variable $X$ given by (\ref{eq:Dust X equations of motion - PG gauge}) is too complicated to be solved exactly.
However, the solution to this variable is not relevant for most purposes, including ours.
We will only need an approximate solution near the maximum-cruvature surface discussed below.

Using (\ref{eq:Structure function - modified - spherical}) to compute the structure function, we obtain the star-interior spacetime metric,
\begin{equation}
    {\rm d} s^2
    =
    - {\rm d} t_{\text{GP}}^2
    + \sec^2 (\lambda K_\varphi) \left( {\rm d} x
    - \sqrt{1 + \lambda^2} \frac{\sin \left(2 \lambda K_{\varphi}\right)}{2 \lambda} {\rm d} t_{\text{GP}} \right)^2
    + x^2 {\rm d} \Omega
    \ ,
    \label{eq:Star interior metric - GP}
\end{equation}
where the $K_\varphi$ expressions are obtained from inverting
\begin{equation}
    \sin^2 (\lambda K_\varphi)
    =
    \frac{\lambda^2}{1 + \lambda^2} \frac{2 m}{x}
    \ ,
    \label{eq:Curvature-mass relation - holonomy}
\end{equation}
with mass function (\ref{eq:Mass function - GP}).
Taking the range $K_\varphi \in ( 0 , - \pi / 2 \lambda)$\textemdash corresponding to $t_{\text{GP}} < 0$\textemdash the signs $s = s_2 =-1$ of (\ref{eq:Interior ansatz - dust - holonomy effects - GP}) and (\ref{eq:Mass equation - holonomy collapse}) are recovered and thus it describes the collapsing region of the star.

Two regions are of interest: near the $x=0$ axis and the $t_{\rm GP} = 0$ axis.
To analyze them we perform a Taylor expansion around $x / t_{\rm GP} =0$ for the former,
\begin{equation}
    \frac{2 m}{x}
    =
    \frac{4}{9} \left(\frac{x}{t_{\rm GP}}\right)^2
    + \frac{16}{27} \frac{\lambda^2}{1+\lambda^2} \left(\frac{x}{t_{\rm GP}}\right)^4
    + O\left(\frac{x}{t_{\rm GP}}\right)^6
    ,
\end{equation}
and an expansion around $t_{\rm GP}/x = 0$ for the latter,
\begin{equation}
    \frac{2 m}{x}
    =
    \frac{1 + \lambda^2}{\lambda^2}
    \left( 1
    - \frac{1}{4} \frac{1+\lambda^2}{\lambda^2} \left(\frac{t_{\rm G P}}{x}\right)^2
    - \frac{1}{48} \frac{(1+\lambda^2)^2}{\lambda^4} \left(\frac{t_{\rm G P}}{x}\right)^4
    \right)
    + O\left(\frac{t_{\rm G P}}{x}\right)^6
    ,
    \label{eq:Perturbation around t=0 - Holonomy collapse}
\end{equation}
This shows that in the limit $x \to 0$, with non-zero $t_{\rm GP}$, the mass function vanishes and one obtains a flat spacetime.
On the other hand, the spacetime metric diverges in the limit $t_{\rm GP}/x \to 0$, thus, this coordinate chart is strictly valid only for the region $(t_{\text{GP}}<0 , 0 < x < R(t_{\text{GP}}) )$.
Using (\ref{eq:Curvature-mass relation - holonomy}), we find that this divergent surface corresponds precisely to a maximum-curvature surface.
As explained in Section~\ref{sec:Spherically symmetric sector}, this is also a reflection-symmetry surface.
Thus, the range $K_\varphi \in ( - \pi / 2 \lambda , - \pi / \lambda)$\textemdash corresponding to $t_{\text{GP}} > 0$ and the sign $s_2=-1$\textemdash, which is a solution too, describes the time-reversed process of an exploding star.

\subsubsection{Near maximum-curvature surface and causal structure}

We now analyze the causal structure.
Null rays follow ${\rm d} s^2 = 0$, or
\begin{equation}
    \frac{{\rm d} x}{{\rm d} t_{\rm G P}} \bigg|_{\text{null}}
    =
    s_{(i)} |\cos (\lambda K_\varphi)|
    + \sqrt{1 + \lambda^2} \frac{\sin \left(2 \lambda K_{\varphi}\right)}{2 \lambda}
    ,
    \label{Causal structure - velocity of null rays - GP}
\end{equation}
where $s_{\rm (in)}=-1$ for ingoing rays and $s_{\rm (out)}=+1$ for outgoing ones.
The causal structure then changes at $x = 2 m$ (that is, $\sin \left(\lambda K_{\varphi}\right) = - \lambda / \sqrt{1 + \lambda^2}$), just as in the classical case, defining a causal horizon where the outgoing null rays become ingoing.
While exactly solving the line $t_{\rm GP} (x)$ for the causal horizon is too complicated, near the maximum-curvature surface we can use the expansion (\ref{eq:Perturbation around t=0 - Holonomy collapse}), such that the causal horizon is approximately given by the line
\begin{equation}
    t_{\rm G P}
    =
    - 2 \frac{\lambda}{1 + \lambda^2} x
    \ .
\end{equation}

There are two important characteristics of the maximum-curvature surface given by $K_\varphi \to - \pi / (2 \lambda)$\textemdash the would-be classical singularity.
First, using the expansion (\ref{eq:Perturbation around t=0 - Holonomy collapse}), we find that the velocity of both ingoing and outgoing null rays do not diverge at this surface, but vanish;
second, the acceleration of null rays remains finite,
\begin{equation}
    \frac{{\rm d}^2 x}{{\rm d} t_{\rm G P}^2} \bigg|_{\rm null}
    \to
    \frac{1}{2 x} \frac{\sqrt{1+\lambda^2}}{\lambda} \left(\pm 1
    + \frac{\sqrt{1 + \lambda^2}}{\lambda} \right)
    \ ,
    \label{Causal structure - acceleration of null rays - maximum curvature surface - GP}
\end{equation}
and positive since $\lambda \ll 1$.

While the previous analysis of null rays suggests that this surface can be crossed, in this expansion the Ricci scalar is given by
\begin{equation}
    {\cal R} =
    \frac{4}{t_{\rm GP}^2} \left( 1
    + \frac{1 + 13 \lambda^2}{24 \lambda^2} \left(\frac{t_{\rm GP}}{x}\right)^2
    - \frac{\left(8-\lambda^2\right) \left(1+\lambda^2\right)}{72 \lambda^4} \left(\frac{t_{\rm GP}}{x}\right)^4
    + O \left(\frac{t_{\rm GP}}{x}\right)^6\right)
    \ .
\end{equation}
Therefore, the singularity at the maximum-curvature surface is a true geometric singularity and no coordinate transformation can remove it.

Finally, using the expansion (\ref{eq:Perturbation around t=0 - Holonomy collapse}), we can solve (\ref{eq:Dust X equations of motion - PG gauge}), to the lowest order to obtain the remaining dust variable,
\begin{eqnarray}
    X &\approx& f_X (x)
    \ ,
\end{eqnarray}
with arbitrary $f_X$, which can be fixed by demanding compatibility (replacing boundary conditions) with the local spatial coordinate, $f_X = x$.

\begin{figure}[h]
    \centering
    \includegraphics[width=0.5\columnwidth]{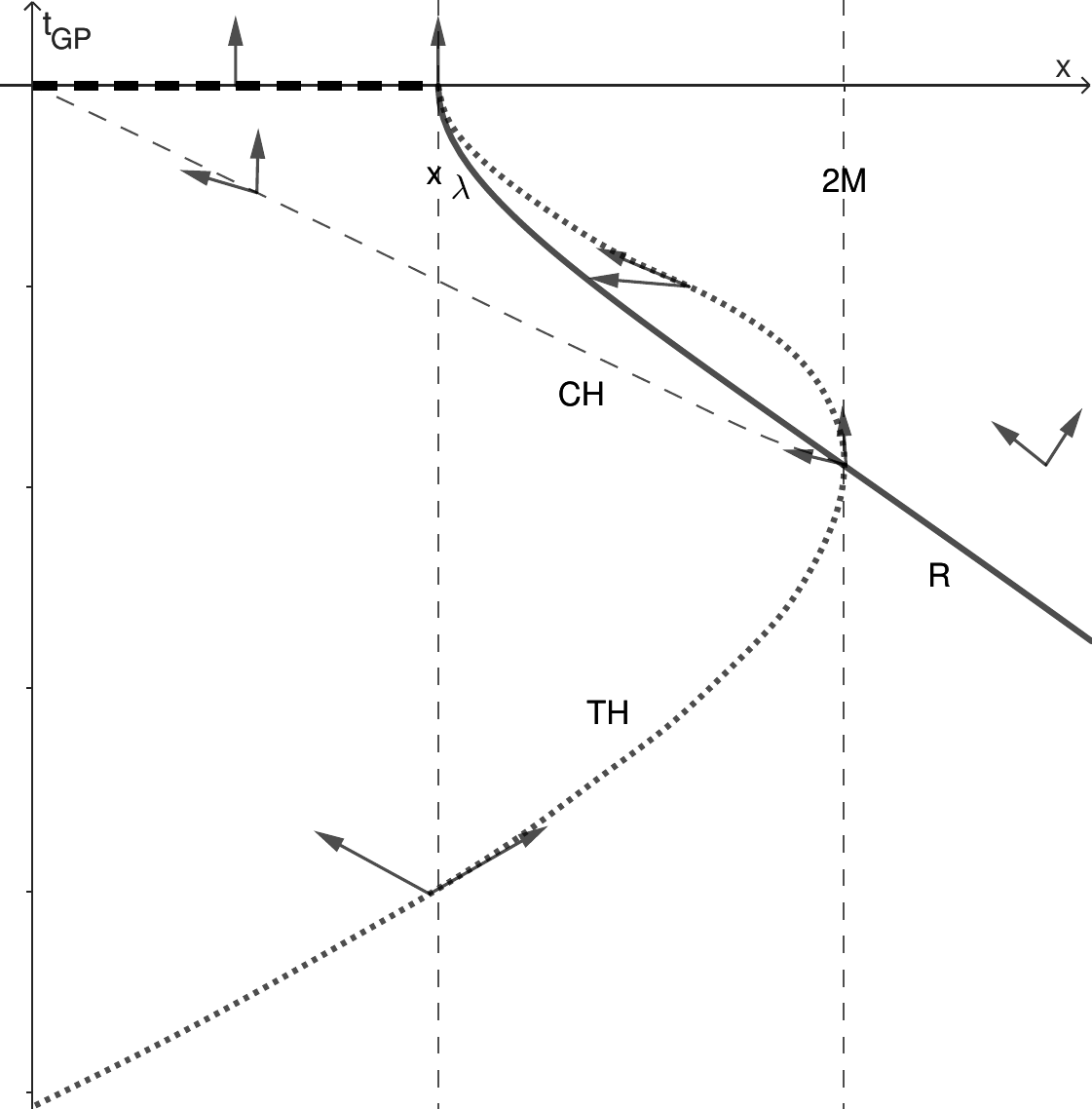}
    \caption{Holonomy collapse of a dust star with infinite initial radius $R_0 \to \infty$ in GP coordinates.
    The star radius $R$ crosses its own horizon at the time $t_{\rm GP} = - \frac{4 M}{3} \left(1+2 \lambda^2/(1+\lambda^2)\right) / \sqrt{1+\lambda^2}$.
    Unlike the classical case, the star's radius does not collapse all the way to $x=0$, but rather to $x_\lambda$ with vanishing velocity and finite positive acceleration.
    Similarly for the trapping horizon $TH$.
    The spacelike singularity of the spacetime lies on the surface $(t_{\rm GP}=0 , x \le x_\lambda)$.
    Some light cones have been represented with arrows for reference on the causal structure.}
    \label{fig:Holonomy_collapse_GP}
\end{figure}

The spacetime diagram of the resulting collapse with holonomy effects in the GP coordinate system developed above is shown in Fig.~\ref{fig:Holonomy_collapse_GP}.


\subsubsection{Non-singular canonical dynamics}

\begin{figure}[h]
    \centering
    \includegraphics[width=0.5\columnwidth]{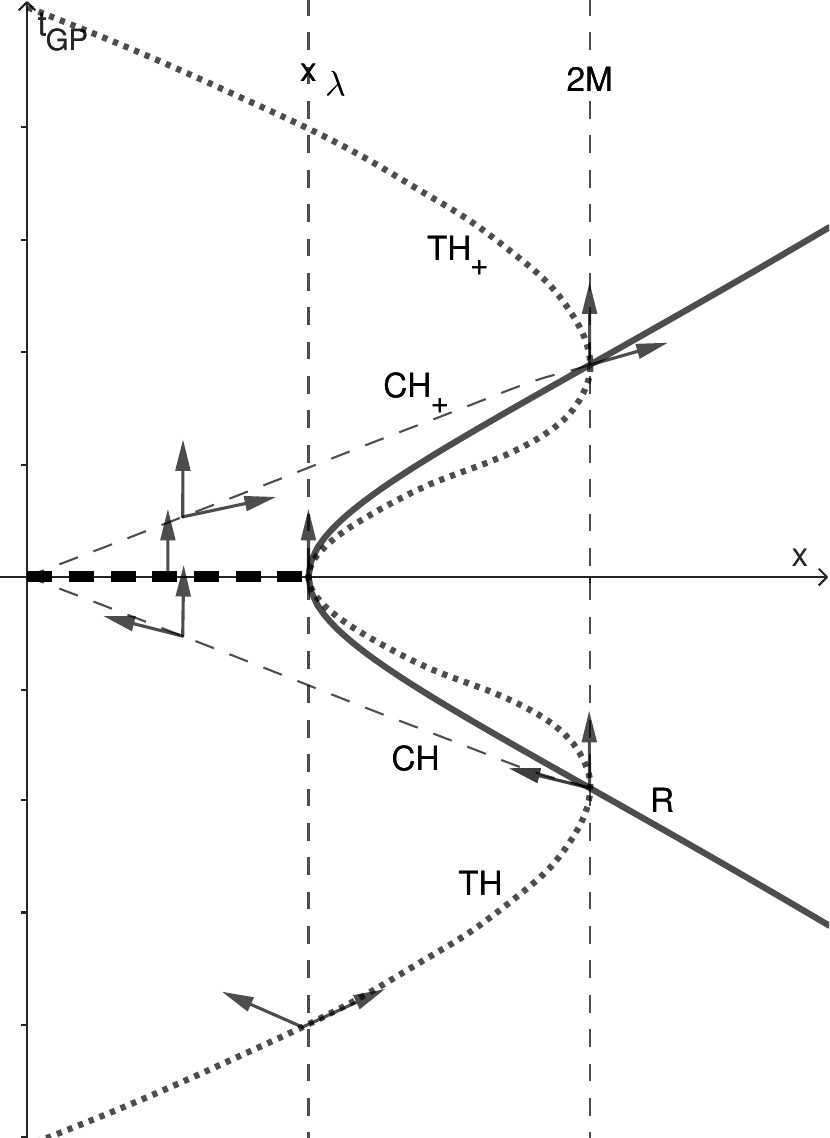}
    \caption{Extension of the star-interior holonomy collapse in GP coordinates.
    The dashed line $(t_{\rm GP} = 0, x<x_\lambda)$ is the maximum-curvature surface containing the spacetime singularity.
    The arrows represent light cones for reference on the causal structure.
    The surface $TH_+$ is an anti-trapping horizon.
    The surface $CH_+$, is an anti-causal horizon beyond which the ingoing null ray's velocity becomes negative once again, and the constant $x$ curves become timelike again.}
    \label{fig:Black-to-White_Hole_Transition_Interior}
\end{figure}

While at the maximum-curvature surface the Ricci scalar diverges, implying a true geometric singularity, the canonical formulation of EMG does not assume the spacetime as a fundamental entity, it is rather an emergent one and, furthermore, it is not the spacetime, but the structure function, the lapse, and shift, that are canonically defined.
Indeed, all of the canonical variables are finite on the maximum-curvature surface:
\begin{eqnarray}
    m
    &\to&
    \frac{1 + \lambda^2}{\lambda^2} \frac{x}{2}
    \ ,
    \hspace{0.5cm}
    N^x \to 0
    \ , \nonumber\\
    P_T &\to& \frac{\sqrt{1+\lambda^2}}{\lambda^2} \frac{1}{2}
    \ , \hspace{0.5cm}
    q^{x x} \to 0
    \ .
    \label{eq:Canonical fields on the maximum curvature surface}
\end{eqnarray}
This differs from the classical canonical values at $K_\varphi \to - \infty$ (and necessarily $x\to R \to 0$) where $m \to M$, $N^x\to\infty$, $P_T \to \infty$, and $K_x \to - \infty$.
The non-singular canonical dynamics of the phase-space variables, which does not rely on the spatial metric, allows us to extend the solution past this surface to complete the full range of the curvature $K_\varphi \in (0 , - \pi / \lambda)$, corresponding to the whole $t_{\rm GP} \in (-\infty,\infty)$.
The first term in the null rays' velocity expression (\ref{Causal structure - velocity of null rays - GP}) shows that ingoing (outgoing) rays remain ingoing (outgoing) after crossing the maximum-curvature surface as determined by the sign $s_{(i)}$; however, since the second term differs by a sign when crossing the surface, the causal structure past the maximum-curvature surface is instead that of a white hole.
The resulting spacetime, up to the divergent surface, can then be pictured by the region $0 < x < R$ in Fig.~\ref{fig:Black-to-White_Hole_Transition_Interior}, which describes a collapsing star of dust followed by its explosion after it crosses the maximum-curvature surface.
The star-exterior region $x>R$ of Fig.~\ref{fig:Black-to-White_Hole_Transition_Interior}, however, is not uniquely determined by the canonical dynamics, since the star-interior solution can be glued to a vacuum solution in different ways, leading to different physical phenomena.
We will explore two such possibilities, one corresponding to the formation of a wormhole, and the other to a black-to-white-hole transition.
To this end, we will make use of the Kruskal maximal extension adapted to the modified theory.

\subsubsection{Vacuum maximal extension}

\begin{figure}[h]
    \centering
    \includegraphics[trim=7.25cm 0cm 7.5cm 0cm,clip=true,width=0.75\columnwidth]{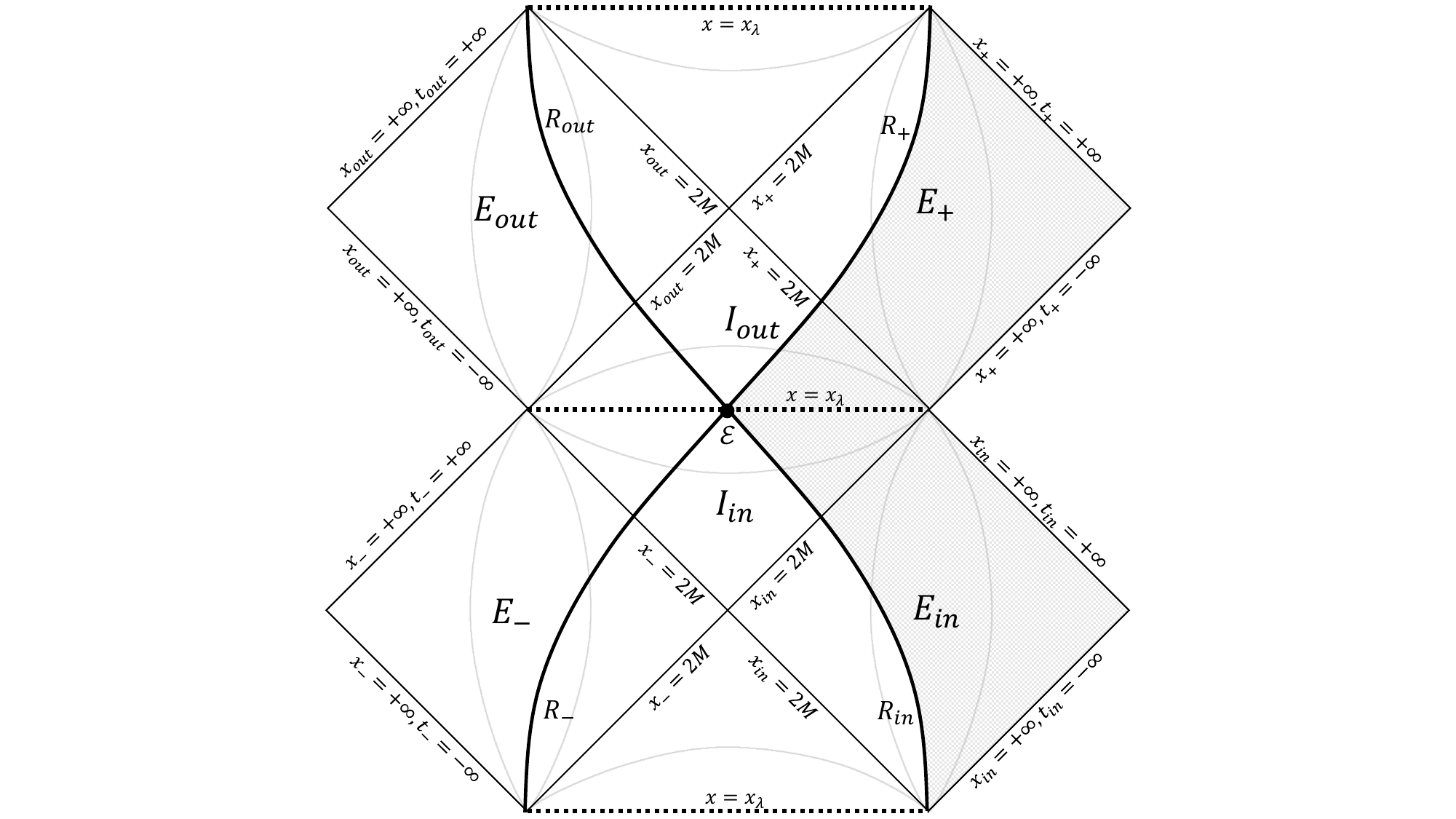}
    \caption{Maximal extension of the wormhole solution in vacuum.
    The faint gray lines denote constant $x$ curves.
    The solid line $R_{in}$ represents a timelike geodesic at rest at spatial inifinity in the remote past of the exterior $E_{in}$. The corresponding geodesics for the other regions are also shown, all intersecting at the point $\mathcal{E}$ on the minimum radius surface.}
    \label{fig:Holonomy_KS_Vacuum_Wormhole}
\end{figure}

Starting with either the modified Schwarzschild metric (\ref{eq:Spacetime metric - Schwarzschild gauge - Static - Holonomy}) or the modified Gullstrand-Painlev\'e metric (\ref{eq:Spacetime metric - GP gauge - Static - Holonomy})\textemdash giving both the exact same result\textemdash we perform the frame transformation to null coordinates (\ref{eq:Kruskal-Szekeres metric form}) to put it in the Kruskal-Szekeres form,
\begin{equation}
    {\rm d} s^2
    =
    - \left( 1 - \frac{2 M}{x} \right) {\rm d} u {\rm d} v
    \ ,
\end{equation}
where the Schwarzschild coordinates are related to the null ones, using (\ref{eq:Null coordinates - Kruskal-Szekeres}), by
\begin{equation}
    u = t - x_*
    \ , \qquad
    v = t + x_*
    \ ,
\end{equation}
with
\begin{eqnarray}
    x_*
    &=&
    2 M \sqrt{1+\lambda^2} \ln \left( c
    \left(\frac{1+\sqrt{1 - x_\lambda/x}}{1-\sqrt{1 - x_\lambda/x}}\right)^{\frac{1 + x_\lambda/(4M)}{\sqrt{1+\lambda^2}}}
    \left|\frac{\sqrt{1+\lambda^2} \sqrt{1 - x_\lambda/x}-1}{\sqrt{1+\lambda^2} \sqrt{1 - x_\lambda/x}+1}\right| 
    e^{\frac{x}{2M} \frac{\sqrt{1-x_\lambda/x}}{\sqrt{1+\lambda^2}}}
    \right)
    \nonumber\\
    &=:& 2 M \sqrt{1+\lambda^2} \ln \Gamma (x)
    \ ,
\end{eqnarray}
where we have absorbed the coefficients $v^{(i)}_t$ into the null coordinates, as we have chosen them to be the constant Killing quantities of null geodesics, and $c$ is a constant.
In the limit $x \to 2 M$, the null coordinates take the values $u \to \infty$ and $v \to - \infty$ as the defined function takes the value $\Gamma \to 0$.
We overcome this by the usual coordinate transformation
\begin{equation}
    U = - e^{-u / ( 4 M \sqrt{1+\lambda^2} )}
    \ , \quad
    V = e^{v / ( 4 M \sqrt{1+\lambda^2} )}
    \ ,
\end{equation}
so that the region $r > 2 M$ corresponds to $U \in ( - \infty , 0)$, $V \in ( 0 , \infty )$.
The metric becomes
\begin{equation}
    {\rm d} s^2
    =
    - F_{\text{E}} \left(x\right) {\rm d} U {\rm d} V
    \ ,
\end{equation}
with
\begin{equation}
    F_{\text{E}} (x) =
    16 M^2 \left(1+\lambda^2\right) \left( 1 - \frac{2 M}{x} \right) / \Gamma (x)
    \ ,
\end{equation}
and $x = x (U,V)$ given by the solution to
\begin{equation}
    U V = - \Gamma (x)
    \ .
\end{equation}
In the limit $x\to 2 M$ we obtain $U , V \to 0$ and finite $F_{\rm E}:$
\begin{equation}
    F_{\rm E} \xrightarrow[x \to 2 M]{\empty}
    \frac{64 M^2}{c} \frac{1+\lambda^2}{\lambda^2} e^{-1/(1+\lambda^2)} \left(\frac{\sqrt{1+\lambda^2}-1}{\sqrt{1+\lambda^2}+1}\right)^{\left(1+\frac{1}{2}\frac{\lambda^2}{1+\lambda^2}\right)/\sqrt{1+\lambda^2}}
    \ .
\end{equation}    
Therefore, we can extend the null coordinates to $x<2M$ by taking negative values for $U , V$, achieving the modified Kruskal extension.
Finally, performing the conformal transformation
\begin{equation}
    \bar{U} = \text{arctan} (U)
    \ , \hspace{1cm}
    \bar{V} = \text{arctan} (V)
    \ ,
\end{equation}
we obtain the metric
\begin{equation}
    {\rm d} s^2
    =
    - \bar{F}_{\text{E}} \left(x\right) {\rm d} \bar{U} {\rm d} \bar{V}
    \ ,
    \label{eq:KS vacuum metric - conformal}
\end{equation}
and the usual Penrose diagram follows, see Fig.~\ref{fig:Holonomy_KS_Vacuum_Wormhole}, which was first obtained in \cite{alonso2022nonsingular}.
Unlike the classical solution, this one is singularity-free in the vacuum.
The maximum-curvature surface at $x = x_\lambda$ is regular and can be crossed.
This surface is one of reflection-symmetry as explained in Section~\ref{sec:Spherically symmetric sector}, thus, on the other side of $x_\lambda$ we simply have the reflected solution of the black hole with the only difference being the flow of time, which gives this region the properties of a white hole.
The global structure is then that of an interuniversal wormhole connecting two different asymptotic regions belonging to the exteriors $E_{in}$ and $E_{out}$ via an interior $I = I_{in} \cup I_{out}$.
Furthermore, the maximal extension allows for the periodic behaviour seen in Fig.~\ref{fig:Holonomy_KS_Vacuum_Wormhole} such that not only two, but four asymptotic regions can be joined by the same interior $I$ of the wormhole.
The curve $R_{in}$ (or $R_{-}$) is precisely the trajectory that the star's radius follows in its collapse stage, while $R_{out}$ (or $R_{+}$) is the trajectory it follows in its exploding stage.

\section{On the possible physical outcomes of the collapse}
\label{sec:On the possible physical outcomes of the collapse}

\subsection{Formation of an interuniversal wormhole}

\begin{figure}
    \centering
    \includegraphics[trim=13.25cm 1cm 12.5cm 1.25cm,clip=true,width=0.5\columnwidth]{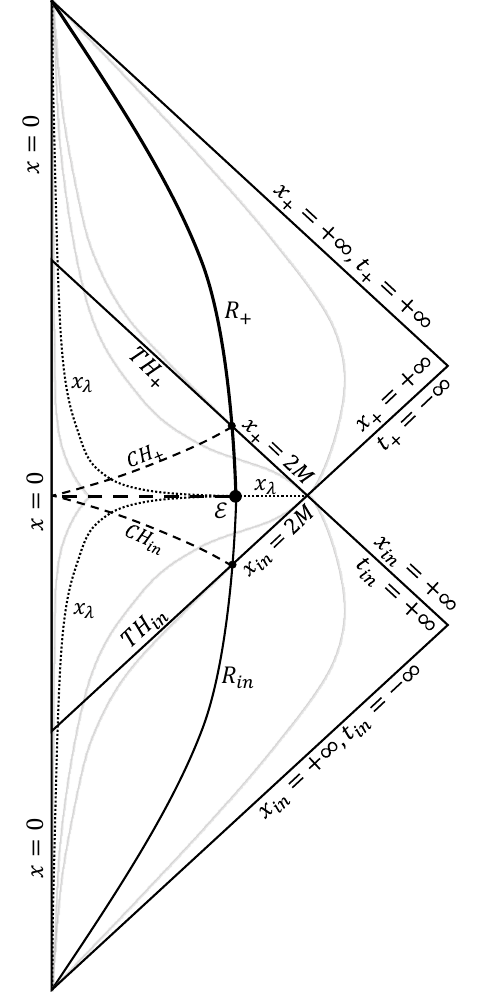}
    \caption{Formation of a wormhole from the collapse.
    For reference, the faint gray lines represent constant $x$ curves, except $x=x_\lambda$ which is the dotted line.
    The dashed line at the center is the $t_{\rm GP}=0$ maximum-curvature surface.
    The curve $R = R_{in} \cup R_+$ is the full trajectory of the star's radius.}
    \label{fig:Holonomy_KS_Formation_Wormhole}
\end{figure}

Consider the vacuum solution shaded in gray in Fig.~\ref{fig:Holonomy_KS_Vacuum_Wormhole} corresponding to the region $E_{in} \cup I \cup E_{out}$ restricted to $x_{in}> R_{in}$ and $x_{+}>R_+$, with boundary $R=R_{in}\cup R_+$.
One can then continuously glue this solution to that of the dust collapse, given by the star-interior of Fig.~\ref{fig:Black-to-White_Hole_Transition_Interior}, by their mutual boundary $R$.
The resulting spacetime can then be represented in a conformal diagram, see Fig.~\ref{fig:Holonomy_KS_Formation_Wormhole}.
The global structure is well-defined everywhere, except at the $t_{\rm GP}=0$ surface joining $x=0$ to $\mathcal{E}$ where the spacetime curvature diverges.
The gluing of the boundary $R$ is everywhere continuous.
This is the spacetime of an interuniversal wormhole joining only two asymptotic regions, rather than the four of the vacuum solution.
The time flow and the causal structure lets an observer in $E_{in}$ to communicate another observer in $E_{+}$ by sending signals that traverse the wormhole, but an observer in $E_{+}$ cannot send a signal back to $E_{in}$.
    
\subsection{Black-to-white-hole transition: Progress in an effective description}

\begin{figure}[h]
    \centering
    \includegraphics[trim=13cm 1cm 13cm 1.25cm,clip=true,width=0.5\columnwidth]{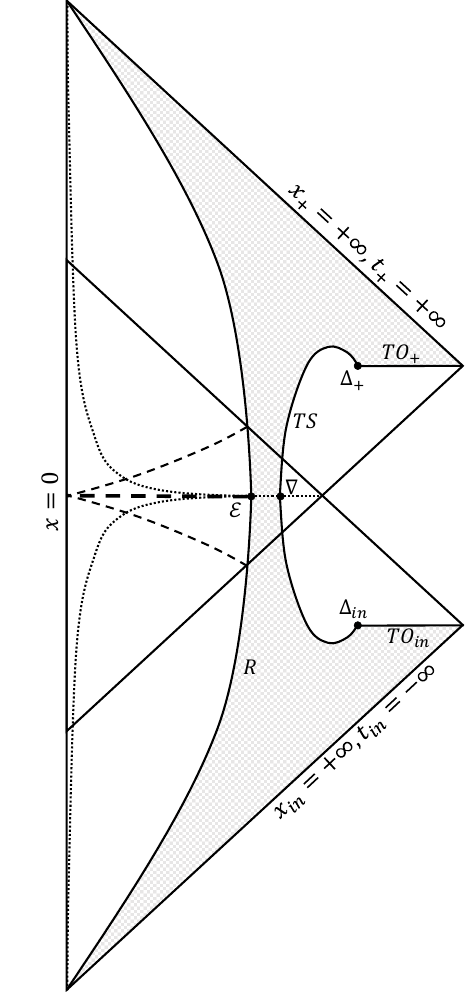}
    \caption{Star-exterior of a black-to-white-hole transition spacetime diagram.
    The shaded region is a solution to the star-exterior modified equations with the two boundaries $R$ and $S = TO_{in} \cup TS \cup TO_{+}$.
    The point where $TS$ intersects the minimum-radius surface is denoted by $\nabla$.}
    \label{fig:Holonomy_KS_BtWH_exterior}
\end{figure}

\begin{figure}[h]
    \centering
    \includegraphics[trim=14cm 2.9cm 13.5cm 3.5cm,clip=true,width=0.5\columnwidth]{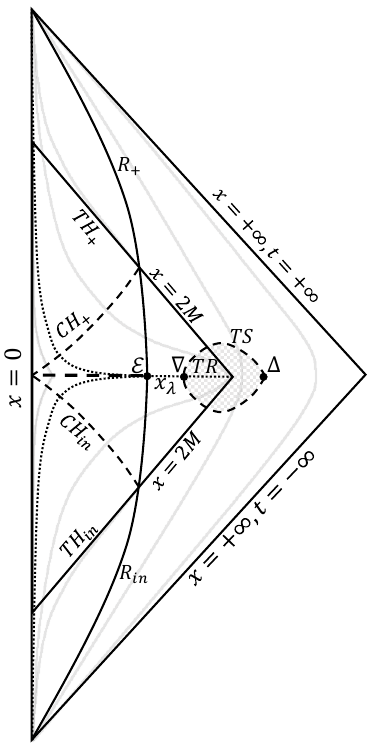}
    \caption{Black-to-White hole transition conformal diagram.
    The dashed line $(t_{\rm GP} = 0, x<x_\lambda$ at the center is the maximum-curvature surface containing the spacetime singularity.
    We refer to the region $TR$ as the transition region, bounded by the surface $TS$.
    The point $\Delta$ is chosen close to the horizon at the coordinates $( t_{\rm GP} = 0 , x = 2 M + \delta)$, with small, positive $\delta$, but otherwise undetermined.
    While the transition region has an undetermined geometry, one can schematically extrapolate the trajectories of constant $x$ curves.}
    \label{fig:Black-to-White_Hole_Transition_KS}
\end{figure}

Following \cite{PhysRevD.92.104020,han2023geometry}, one can obtain a spacetime describing a black-to-white-hole transition by slicing a wormhole spacetime of the form of Fig.~\ref{fig:Holonomy_KS_Formation_Wormhole} such that a single asymptotic region remains.
This slicing can be done as shown in Fig.~\ref{fig:Holonomy_KS_BtWH_exterior}.
The shaded area is a valid solution to the vacuum equations of motion with two boundaries, one given by the star's radius $R$, and the other given by $S$.
The two surfaces $TO_+$ and $TO_{in}$ can then be smoothly joined because the spacetime surrounding them is locally identical.
The resulting conformal diagram is shown in Fig.~\ref{fig:Black-to-White_Hole_Transition_KS}.

The details on the surface $TS$ and the precise positions of the points $\nabla$ and $\Delta$ are not provided by the canonical equations of EMG, and must be provided by considerations of a particular quantum gravity approach (or whatever underlying theory is being modelled), so we leave them as undetermined in our effective approach\textemdash for instance, the construction of \cite{PhysRevD.92.104020} estimates the coordinates of the point $\Delta$ to be $(t=0 , x = 7 M / 6)$ in classical Schwarzschild coordinates.

The effective theory thus provides a picture compatible with the black-to-white-hole transition proposal as a quantum gravity phenomenon in this particular gluing of the dust and vacuum regions.
However, the region $TR$ bounded by the surface $TS$ is still unsolved, and more work is needed to determine whether it can be described by EMG as the effective theory or if more details of quantum gravity are needed.

If we take the divergence on the maximum-curvature surface as a sign of the break down of EMG as an effective theory, then we can insert a second transition region surrounding the surface joining $x=0$ to $\mathcal{E}$, or simply extend the single transition region to surround the maximum-curvature surface by placing $\nabla$ at $x=0$.


\section{Conclusions}
\label{sec:Conclusion}

We have outlined a procedure to couple matter to EMG on purely canonical terms ensuring general covariance of both the spacetime and matter fields.
We have explicitly done so for the perfect fluid, obtaining the general form of the perfect fluid contribution to the constraint, and found that no modifications are allowed except for the use of an emergent metric and a non-trivial gravitational dependence of the pressure function.
Finally, we have also outlined a procedure to study gravitational collapse in spherically symmetric EMG using a comoving coordinate system corresponding to the canonical formulation of the Oppenheimer-Snyder model, and we have explicitly computed the collapse of dust.

The dynamical solution to the gravitational collapse of dust in EMG, modelling certain effects of quantum gravity, predicts that a spacetime singularity is formed at the maximum-curvature surface.
However, despite such singularity, and unlike the classical case, the (ingoing and outgoing) null geodesics acquire a vanishing velocity and positive, finite acceleration at the maximum-curvature surface.
Furthermore, all the canonical variables remain finite too, suggesting that the canonical solution can be extended past this surface and 'bounce'.
Consequently, we have obtained a spacetime metric that describes the entire region below the star's radius.
However, since a consistent gluing of the star-interior solution to the star-exterior one describing the vacuum does not follow uniquely from solving the equations of motion, the physical result of the collapse will depend on the specific choice of the gluing process.

The most straightforward gluing leads to an interuniversal wormhole spacetime with two distinct asymptotic regions that can be crossed in a single direction.
This wormhole connects a nascent black hole as the result of the collapse in the first universe to an evaporating white hole as a result of the matter emerging out of it in the second universe.
The second gluing, which results from slicing out part of the wormhole spacetime, consists of a single a asymptotic region that fits with the black-to-white-hole transition proposal suggested by some quantum gravity approaches.
As a compromise of such slicing, a finite transition region $TR$ becomes undefined which, together with the divergent geometry of the maximum-curvature surface, presents an ongoing challenge to this proposal in an effective description.
The resolution of this transition region might be available only in a full quantum gravity treatment or, at least, with more input to the effective approach.
Instabilities of the black-to-white-hole transition when perturbations are present studied with a classical geometry can be tamed by a time-asymmetric scenario \cite{de2016improved}; thus, future work will be devoted to ensure the EMG model cures these instabilities too.

Though more work is needed to resolve the above issues, and to determine whether they appear in the collapse of different forms of matter, it is encouraging to learn that new properties of gravitational collapse and the ultimate fate of black holes can be found in covariant modified gravity theories such as emergent modified gravity.

\section*{Acknowledgements}

The author is grateful to Martin Bojowald for discussions.

\end{document}